# Multidimensional Cybersecurity Framework for Strategic Foresight

**Cyril Onwubiko[1,2] and Karim Ouazzane[3]**
[1]*Enterprise Security Architecture, Chief Information Security Office (CISO), Pearson Plc*
[2]*Artificial Intelligence, Blockchain & Cyber Security, Research Series, London, UK*
[3]*Cyber Security Centre, London Metropolitan University, London, UK*

**ABSTRACT**
Cybersecurity is now at the forefront of most organisations' digital transformative agendas and National economic, social and political programmes. Hence its impact to society can no longer be seen to be one dimensional. The rise in National cybersecurity laws and regulations is a good indicator of its perceived importance to nations. And the recent awakening for social and ethical transparency in society and coupled with sustainability issues demonstrate the need for a paradigm shift in how cybersecurity discourses can now happen. In response to this shift, a multidimensional cybersecurity framework for strategic foresight underpinned on situational awareness is proposed. The conceptual cybersecurity framework comprising six domains – Physical, Cultural, Economic, Social, Political and Cyber – is discussed. The guiding principles underpinning the framework are outlined, followed by in-depth reflection on the Business, Operational, Technological and Human (BOTH) factors and their implications for strategic foresight for cybersecurity.

## I. Introduction

Cybersecurity is now an enterprise, national, global and societal priority, an utmost priority for that matter. Emerging cybersecurity discourse and debates permeate society, whether it is in guiding against false news and misinformation in social media networks and platforms, or regulating Blockchain to sustainability (Environmental, Social and Governance (ESG)) concerns, or ethical use of Artificial Intelligence (AI) to enacting new National cybersecurity laws, and data privacy directives and regulations. These developments correlate with the increasing number of countries creating their own National cybersecurity strategies (e.g., UK Cybersecurity 2021, US National Cyber 2018, Finland Cybersecurity 2019, Australia Cybersecurity 2020), and regional national cybersecurity centres (e.g., the European National Information Security Agency (ENISA), ASEAN[1] Cyber Capacity Programme, and ASEAN-Japan Cybersecurity Capacity Building Centre etc.). And the growing national and regional data privacy regulatory derivatives (e.g., European GDPR), national cybersecurity laws (e.g., Chinese Cyber Security Law) in part, to protect themselves, their intellectual property from cyber espionage and foreign State interference in the rule of law and governance. In the other part, to foster a secure and safe cyberspace to conduct businesses and to attract human capital, talent pipeline and bolstering innovation and economic potential and opportunities.

Strategic vision for cybersecurity as a business decision attempts to balance protection needs with the desired business outcomes (Proctor and Sampath, 2020), and understanding that it can no longer be solved through technical approaches alone is equally as important. Cybersecurity approaches aimed only at the technical and operational levels have been shown to be inefficient in countering emerging cyber threats (Gandhi R. et al., 2011, Ikwu R. et al., 2017, Nakhla N. et al., 2017), and the best defenses are found to be both technical and organisational (Marotta A., and Pearlson K., 2019). According to David Koh, Chief Executive of the Cyber Security Agency (CSA), "Cybersecurity can no longer be considered merely a *technical issue*, but has spilled over to the *geopolitical realm.* Critical technologies and communications infrastructure, to emergent ones such as 5G, cloud computing, artificial intelligence, and quantum computing, and the associated cyber risks they portend, provide conduits to impact a country's National security, economic progress, and societal values." (Koh, 2020). In recognition that cybersecurity is now a societal priority, the United Nations agreed *11 consensus norms of responsible State behaviour in cyberspace*, articulated in the 2015 U.N. Group of Governmental Experts (GGE) report (U.N. GGE, 2015).

---
[1] ASEAN – Association of Southeast Asian Nations

It is in response to these current debates and realities that we propose a multidimensional cybersecurity framework to extend the realms of coverage to include Physical, Cultural, Economic, Social, Political and Cyber dimensions.

The contributions of this paper are:
1. We propose and explain a multidimensional cybersecurity framework comprising six domains, namely Physical, Cultural, Economic, Social, Political and Cyber.
2. We outline the foundational principles underpinning our multidimensional framework, and through a derivative of Business, Operational, Technological and Human (BOTH) factors we reflect on strategic foresight in cybersecurity through these viewpoints.

The remainder of this paper is organised as follows: Section II discusses motivation for the research and related contributions in the literature. Key terminologies used in the paper are defined, and methodologies applied in the work are outlined. Section III explains our proposed multidimensional cybersecurity framework. It discusses the key principles applied in the research, and provides an in-depth discussion of the Business, Operational, Technological and Human factors. Section IV compares intra-domain and inter-domain cybersecurity situational awareness and their relevance; finally, conclusions and recommendations for future work are presented in Section V.

## II. Background

### A. Motivation

The purpose of a National or organisational cyber programme is to design, build and embed a robust cybersecurity practice for a nation or an organisation that allows its citizens or stakeholders to conduct business in a secure and safe manner that bodes trust for the nation or organisation and its stakeholders in the ecosystem (i.e., citizens, allies, clients, partners and supply chain). To achieve this purpose, the cyber programme must understand the business operatives of the nation or organisation, the risks it bears, the threats it faces, and most importantly, the business requirements and responsibilities of the nation or organisation. This is very much so that the cybersecurity capabilities being deployed by the programme are relevant, efficient and achieves the set goals. With the many objectives that an organisation may have, its overarching cybersecurity goal is to ensure that <u>D</u>ata, <u>A</u>pplications and business and information <u>A</u>ssets and <u>S</u>ervices (DAAS) of the organisation are appropriately protected. It should also have moral, ethical, and socio-economic (and socio-political for a government or National cybersecurity programme) responsibilities of the security of the products and services it offers, and the supply chain and partners in which it conducts businesses with. Finally, it should have sustainability (environmental, social and governance) responsibility of the impacts its products and services have to society at large.

To some organisations, protection is focused on procuring expensive technical and operational cyber defence tools. We argue that this approach does not offer comprehensive protection needed for emerging cyber threats and attacks. According to several contributions e.g. (Ikwu R., 2017, Gandhi R. et al., 2011), cyber defense approaches centered only around technical and operational dimensions are inefficient, incapable and limited.

Our approach to cyber defense is holistic and multidimensional encompassing not only technical and operational, but also social, cultural, political and physical dimensions. This means ensuring that there are trained and capable *people*, robust *processes, social* and *technical capabilities* to identify, prevent, detect, monitor, respond and recover from cyber-attacks and cyber incidents. Economic and political dimensions such as geo-political, regulatory and sanctions-based controls to deter and enforce adherence and compliance. Furthermore, we emphasise on the concerted efforts for the technical, socio-cultural and human capabilities to be integrated and responsive. Economic and political controls to be coordinated, while technical and operational capabilities to be automated. Our proposal for human-in-the-loop for cybersecurity management and decision making is to provide strategic cyber-foresight.

### B. Related Work

The seminal work of Mica Endsley on situational awareness (Endsley, 1995) is regarded as the pivotal point for this field of study, and since then, situational awareness has been applied to several areas e.g., safety, security and transportation. Situational awareness has also been applied in cybersecurity, vehicular networks, aviation and social media analytics (Onwubiko C., 2016, Eiza M. H., 2017). Situational awareness is ideal for understanding organisational and human factors aspects that offer insights on human operator decision making (e.g., cognition, teamwork, knowledge, skills and abilities). We see this to be pertinent in this paper, especially in gaining 'understanding' of the relationships through humans-in-the-loop, who understand and have experience of cybersecurity assessment, major incident management, geo-politics, cyber economics and cyber physical systems. These humans-in-the-loop leverage technology, automation and integration combined with their experience, skills and knowledge to gain cyber foresight. Furthermore, the interdependence and inter-dimensionality of the multiple domains e.g., physical, cultural, economic, social, political and cyber that should be considered in order that enhanced situational awareness across the domains can be achieved.

Cyber defence tools are not a 'silver bullet', and do not solve all the cybersecurity problems themselves. For example, cyber defence tools such as firewalls or intrusion detection systems are unable to solve cybersecurity procedural or human factors problems. They are as efficient as the people who use them to monitor business services, follow up on incidents and conduct incident triage. The tools may offer cues and prompts which the human operators, such as Cyber Security Operations Centre (CSOC) administrators and analysts should investigate. Often these cues are *symptomatic* - an expression of a likelihood of something, rather than an explicit indication, therefore, human expertise and experience are very much required. The cues which are provided by the monitoring systems may be in the form of alerts, alarms, events etc. These intelligences will need to be analysed and decision on possible cause of action (CoA) will be down to humans to make.

A cross-domain CyberSA framework was proposed by (Hall M. J., Hansen D. D., and Jones K., 2015), which focused on data, information and intelligence but within a single domain (Cyber Domain). Their concept of domain relates to data, information, knowledge, intelligence. Similarly, the Australian Government's Cyber Security Centre (Australian ACSC, 2015) provides guidance for Cross-Domain solution on how to connect systems of different business impact levels, and security classifications e.g., OFFICIAL, SECRET & TOP SECRET. Equally, the focus of that guidance was specifically on impact levels rather than domains in the sense of different realms.

We argue that data, information, knowledge and intelligence are 'layers' as opposed to domains. Our use of *domain* in this paper relates to P*hysical, Cultural, Economic, Social, Political* and C*yber*. These are distinct realms that are governed by specific laws, attributes, and intrinsically and extrinsically by unique features, like the traditional domains, for example, Land, Sea, Air and Space. We refer to data, information, knowledge and intelligence as *layers* that exist within a single domain. Data can be collected from the physical, cultural, economic, social, political or cyber domains. For example:
- Data relating to the physical domain, may be temperature, humidity and the HVAC (heating, ventilating, and air conditioning) of the data centre, environment or location.
- Data relating to cultural domain may be human actions, such as decisions made by operators of the systems, their reaction to a cyber incident and how quickly they invoked incident response playbook.
- Data relating to economic domain may be related to business models about business outcome or stock market in the event of a negative publicity resulting from cyber incident or security breach.
- Data relating to social domain may be related to diversity, equity and inclusion of the workforce e.g., operators of the systems. The more diverse they are, the better their collective intelligence, and hence better response approach and playbook.
- Data relating to political domain may be how citizens and corporations react to a new piece of legislation or policy, or how a sovereign nation reacts to when it finds out that a recent cyber incident it suffered was conducted by a hostile nation.
- Data relating to cyber domain may be related to cyber incidents to digital and online infrastructures.

It is pertinent to note that the six (6) domains are interdependent and interwoven (See fig. 1). Take for example, the Physical Domain and the Cyber Domain are interdependent, and their intersection has borne a class of new systems referred to as the *Cyber Physical Systems* (CPS). Similarly, there are socioeconomic, geopolitical, sociopolitical and sociocultural systems, and many others, this is just to show the various interdependencies across the domains.

### C. Definitions

*Cybersecurity* is the protection of information, data, communications and the underlying infrastructures that are used to process, store and transit information and data. It is about ensuring confidentiality and integrity of the information and data it processes and/or stores, and the availability thereof to the appropriate/legitimate owners and users of the system. It has been defined in several complementary ways, such as an enabler for the advancement and growth of economic and social prosperity, see (Finnish Cyber Security Strategy, 2010, UK Cyber Security Strategy, 2021, NATO CCD COE, 2015).

*Cyber defence* is defined with respect to offensive and defensive capabilities, that is, the ability to defend, stop and counter cyber-attacks and threats through cyber means for resilience, recovery and business continuity. Cyber defense is more akin to military and agency cyber programme where the capability to *prevent* has more emphasis than *detection* and *response*. It is important to note that these terms are occasionally used interchangeably in the literature, (see, DoD MNE7, 2013).

According to (Dictionary Online, 2019), *foresight* is defined as "*knowledge* or *insight* gained by or as by looking forward; a view of the *future*". Similarly, the (Oxford Dictionary, 2019), defines *foresight* as "the ability to *predict* what will happen or be needed in the *future*". Based on these definitions of foresight, we define *Cyber foresight* (a.k.a. Cybersecurity foresight) as the ability to understand current cyber situations or circumstances in order that we may be able to predict and address future or emerging cybersecurity situations. By this definition, cyber foresight encompasses knowledge, understanding, or insight into the current cyber situations or circumstances with a view to predicting future cyber situations. *Cyber situations* include cyber threats, cybersecurity attacks, cyber risks and cyber issues, such as vulnerabilities, exploits, security incidents, data breaches and cybercrime.

*Cyber Situational Awareness (Cyber SA)* has been defined in many ways in relation to cybersecurity, cyber defense, and cyber operations (see, Tadda G. P., and Salerno J. S., 2010, Cumiford D. L., 2006). We adopt the definition of Cyber SA provided by Onwubiko and Owens (Onwubiko C. & Owens J.T., 2011), which states that "situational awareness is the ability of being aware of circumstances that exist around us, especially those that are particularly relevant to us and which we are interested about. It encompasses the prediction of future states or impending states because of the knowledge (which could include new information) and understanding of current states".

Based on these definitions, one would argue that there exists an overlap between Cyber Foresight and Cyber SA. Also, we believe that both are congruent and complementary. This is because both seeks understanding and insight with the ability to predict. It would not be surprising if 'cyber foresight' and 'cyber situational awareness' are used interchangeably in certain contexts and application.

### D. Research Methodology

The following research methods are applied in this paper to investigate multidimensional situation-aware security framework:
- *Attribute Listing:* We use attribute listing to enumerate the domains considered for the proposed multidimensional framework (see Section III and Fig. 3). Attribute listing is a research and innovative methodology formulated in the early 1930s that allows all possible attributes/features of an object or a thing to be listed; including smaller aspects of the object that would ordinarily not been considered in conventional listing. By doing so, it allows for innovative understanding and discovery of some aspects of the existing object to be revealed or uncovered which would ordinarily would have been unknown, hence the newly discovered attributes can then be combined in new ways that were not known prior to develop new products, gain a greater understanding of existing objects or to innovate and extend the functionality of existing objects.
- *PEST Analysis:* We use PEST (Political, Economic, Social and Technological) Analysis to conduct impact assessment of the Business Factors of the proposed framework as discussed in Section III C3 (see Table 1) of the paper.

- *Human Factors:* We use Human Factors Analysis and Requirements Elicitation Method (see Section III B) to investigate Human Factors of the framework, which is discussed in Section III C4 of this paper.
- *Theory of Work Adjustments (TWA):* We apply and discuss in Section III C4 in relation to Human Factors and the SOC analysts' operational needs and how to ensure Human Factor Blockers such as stress, anxiety and workload are minimised. TWA argues that individuals and environment are mutually satisfied if and only if the Knowledge, Skills and Abilities (KSA) of the individual meet the demands of the environment (Dawis and Lofquist, 1987).
- *Training Needs Assessment (TNA):* TNA is another methodology used in the study to ensure that appropriate training is provided to cyber analysts to do their work, this is covered in Section III C4 under Human Factors.
- *Fieldwork and Experiential Analysis:* Our experience and fieldwork (case studies) offer contributory insights to this study.

### III. Multidimensional Cybersecurity Framework

A Multidimensional Situation-Aware Cybersecurity Framework is proposed (See Fig. 1). It consists of four key components namely 1) D*omains* (Physical, Cultural, Economic, Social, Political and Cyber), 2) F*actors (Business Operational, Technological and Human)*, 3) C*ross-Domain Situational Awareness* and 4) C*yber Foresight*. Our framework allows Cyber Foresight to be gained through the understanding of the relationships that exist across the various Domains combined with consideration of the impacts, causes and effects of the concepts identified across the Factors – Business, Operational, Technological and Human (BOTH) factors. Cyber foresight is the culmination of the understanding and comprehension of the causes and effects, impacts and influences of the factors and their relationships across the domains.

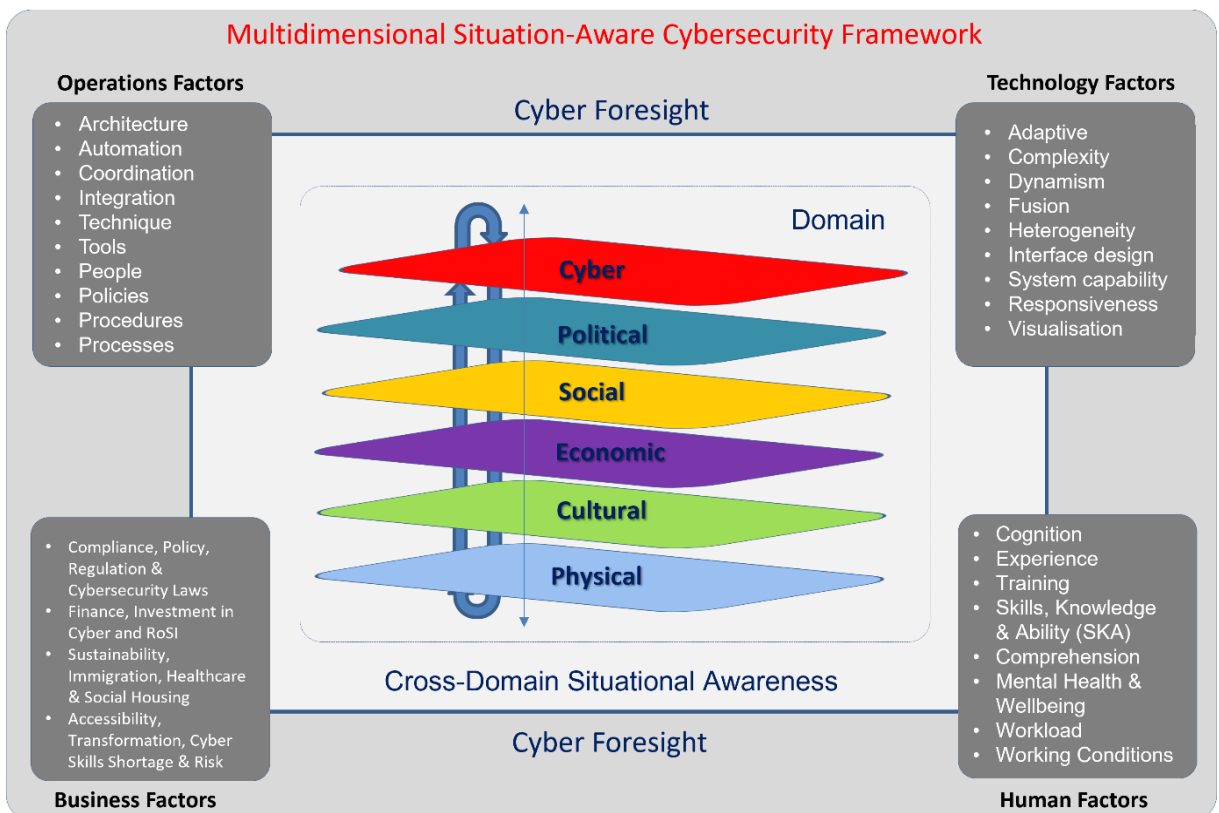

Figure 1: Multidimensional Situation-Aware Security Framework

The framework is composed of:
- Six (6) Domains: Physical, Cultural, Economic, Social, Political and Cyber. These domains are selected through several iterations of Attribute Listing methodology. As discussed in the related work section. This work is novel and unique in its consideration of six domains, and there is no research, to the best of our knowledge that has considered the fusion of these domains in the way and manner applied in this research.
- Four (4) Factors: Business factors, Operations factors, Technology factors and Human factors. Each factor contains a list of attributes and features as shown in Fig. 1. The justification for the Factors and their features is selected through a combination of A*ttribute Listing* and *PEST* methodologies, which are explained in the subsections.
- Cross-Domain Situational Awareness: This entails the awareness and comprehension of the issues arising from the six domains and using this information to find the best resolution to the problems, and in addition, potentially, being able to use them to make certain predictions of future states.
- Cyber Foresight: This is effectively the 'outcome' of combining the information from the domains, and consideration of the four factors and their constituent subfactors into gaining enhanced situational awareness and informed decision making.

### A. Domain

In this paper, we describe a Domain as a **realm** or **dimension**, which is governed by a set of well-defined rules and principles that control its operation and use. **Domain** is used in this paper also as a way of explaining the dimensions or realms that impact strategic cyber programme or mission. And when these domains are not considered they could be exploited to affect the mission's capability to protect itself and/or its constituents. The domains of interest for this research and that are considered to gain strategic cyber foresight and superiority for the nation or mission are Physical, Cultural, Economic, Social, Political and Cyber (as shown in Fig. 2).

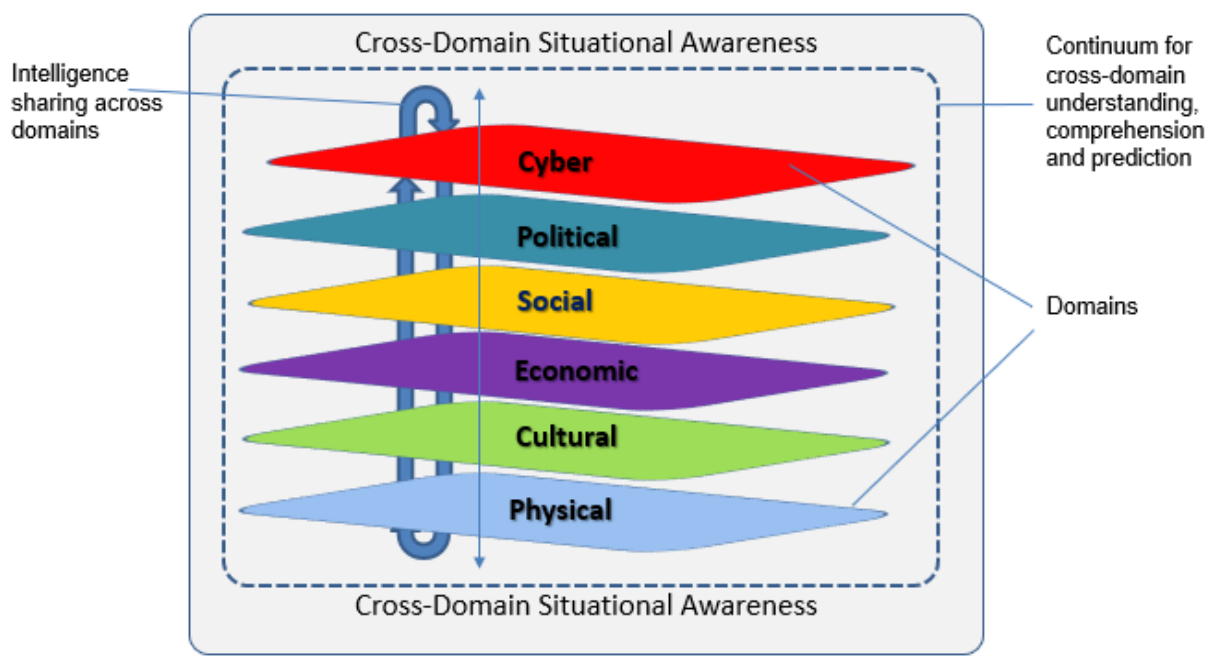

Figure 2: Cross-Domain Situational Awareness

To the best of our knowledge, our proposed framework is by far the most comprehensive framework for S*trategic Cyber Foresight*. We note that (Romanych M., 2010) considered three domains – Physical, Information and Cognitive – for Information Operation in a military context, while (Alberts D. et al.., 2001) considered the same three domains but in relation to Information Age Warfare. These contributions are foundational and fundamental to Cyber Foresight. Since we framework builds on the works of (Romanych M., 2010) and (Alberts D. et al.., 2001) they provide the foundational basis.

For example (as will be discussed below), the Physical Domain in this paper subsumes the well-known *domains* such as Land, Sea, Air, and Space. While the Cyber Domain encompasses the Information, Intelligence, Data and Knowledge domains, which have been studied in their own rights as distinct domains (see, Hall M. J., Hansen D. D. and Jones K., 2015).

The *Physical Domain* deals with the physical 'things' such as the location, geography, environment, buildings, architecture, communications infrastructures, systems, networks, media, and including hosting (building), data centre and environmental facilities, such as water, power (electricity & gas), and wind turbine. Additional examples can be seen in Table 1. In a broader sense, it encompasses Land, Air, Sea and Space. According to the Department of Defence (DoD) Command and Control Research Program (CCRP), the Physical Domain is the real tangible world. In the context of this paper, the Physical Domain comprises the physical infrastructure of an organisation such as building, network architecture, systems, networks and storage, SCADA, IoT and cyber-physical systems. These are the things that the programme must consider and assess when developing a robust solution. Equally, because threats are realised on the physical infrastructures, it is therefore imperative to ensure adequate protection of the Physical Domain.

The *Cultural Domain* deals with people, cognition and metal models (a.k.a. perspective and perception), communities, groups, National and organisational shared values, behaviors and practices (see Table 1). Every organisation has its own organisational culture, e.g., diversity and inclusion culture, learning and development culture, cybersecurity culture, governance culture etc. One could ask, what is the cybersecurity culture of that organisation? Or what is their governance practices and culture? For example, is it a hierarchical culture, are reporting lines based on hierarchy, or is it an open and diverse culture, one that allows people to learn, grow and develop, or is it bureaucratic, one that hinders and stifles innovation and collaboration? Cybersecurity culture is one aspect of organisational culture (Marotta and Pearlson, 2019), but a very important aspect especially in the context of this study. For example, Uchendu et al. (2021) states that the Culture Domain has been heavily studied in the past decade in research and practice as organisations seek to combat increased cyberattacks that focus on exploiting human factors. It is important to recognise that organisations in certain verticals are known to be more regulated in certain respects than others, and hence may require more stringent compliance and enforcement cybersecurity culture. For example, cybersecurity culture in say government departments, military or defence organisations are much more stringent compared to non-defense related organisations. The same can be said of regulatory culture for financial institutions and healthcare organisations. The Cultural Domain is also known as the cognitive domain (see Borgatti S. P., and Halgin D. S., 1998), informed by shared perception, beliefs and values (Romanych M., 2018). The people aspect should be considered and assessed when developing a robust security programme. We rely on people for protecting the enterprise, on compliance with directives, and good judgement and decision making, hence the Culture Domain is of the same importance as the other domains.

The *Economic Domain* deals with the financials, goods, products, services, supply and demand, supply-chain economics and return on investment of cybersecurity (see Table 1). Economics of information security is well studied. Anderson R. and Moore T. (2006) are pioneers of this discipline who argue that 'security failure' is caused at least as often by bad incentives as by bad design. They went on to discuss several challenges in the economics of information security by considering *misaligned incentives* and *externalities*. Felici et al. (2016) postulate that as ICT is used in the cyber domain e.g., information technology and in the other domains, e.g., physical and economic domains, hence, cybersecurity economics is essential in enabling ICT to hold this dual role (i.e., applied to both the cyber and other domains). Hence the economic impact is not just on one domain rather on multiple domains. Evolution in the cyberworld over the last decade has introduced further conundrum in the economics of cybersecurity. For example, 1) recent cyber-economic models such as ransomware as a service (RaaS), where malware infrastructures can be rented on a 'pay as you go' service by bad actors to conduct cyber-attacks (Midler M. (2020)) has introduced further complications into cybersecurity economics. 2) while cyber insurance in turn offers insurance cover over cybersecurity breaches to an organisation, it has introduced a new dimension (i.e., a new economic model) to this field. 3) supply-chain risk inherent in business models, e.g., Business to Business (B2B), Direct to Consumer (D2C) business models and Open Source Software (OSS) models, such as freemium, freeware, open source, community source etc., have introduced varied responsibility issues. For example, who owns the codes, how can it be used, who maintains and has ultimate responsibility for code assurance. Without understanding the supply-chain ecosystem or the goods, products or services of the organisation, how could a robust security programme be designed and implemented? And how can the cybersecurity risks be identified, or cybersecurity situations be recognised, let alone be understood and addressed. To gain enhanced situational awareness, that is, an understanding of current situations, and the ability to make predictions of future states, then the Economic Domain, and factors influencing it should be known, assessed and understood. We are aware of several supply-chain exploitations, and allegations of nation state cyber covet monitoring and intellectual property theft through supply-chain ecosystem (Reed J., 2012, BBC, 2018, Chapman B., 2018). Therefore, we argue that cybersecurity foresight should encompass an understanding of supply-chain ecosystem, and the risks they pose. The Economic Domain should also consider economic incentives e.g., measures to assess cybersecurity effectiveness, and regulatory and liability mandates to better improve security processes (Lesk (2011)). Further, it should consider investments into cybersecurity, the business case for cybersecurity, the economic benefits and benefits realisations, and the return on cybersecurity investment (RoSI). Every organisation, and in fact, the nation or government needs to know their threshold in cybersecurity investment (Brangetto P., and Aubyn M. KS., 2015), and the objectives it should deliver to society, government and citizens; whether *direct tangible benefits* (e.g., having a secure and reliable National cybersecurity strategy could protect citizens and Critical National Infrastructures) and *indirect tangible benefits* (e.g., having secure and reliable National cybersecurity strategy could attract foreign investments and create new job opportunities for citizens alike).

The *Social Domain* is related to the social, ethical, sustainability, environmental, governance and other relational considerations that influence cybersecurity practices of the organisation or the nation (if it is a National cyber programme). Additional examples can be seen in Table 1. We reflect on the relationships, associations and communications involved in managing cybersecurity within an organisation (e.g., intra-cybersecurity practices and management) and with their supply chain, vendors and partners (inter-cybersecurity practices and management). In discussing the social aspects of information security, Frangopoulos et al. (2008) identify three categories of stakeholders in an organisation, *security professionals* (who create and direct information security for the organisation), *management* (who provide oversight for the programme) and *end-users* (who should adhere to information security policies and directives). In examining the relationships and cooperation among these groups, and the difficulty in information security compliance in the organisation, they conclude that "the difficulty in common acceptance and internalisation of the security effort by all stakeholders of an organisation creates innumerable security holes and provides social engineers with the opportunity for successful attacks". Extrapolating Frangopoulos et al.'s argument, one could not but imagine the extremely challenging situation it could be when dealing with supply chain ecosystem. Trust in social, digital and online enterprises is key. How can we trust that stakeholders in an organisation (security professionals, management and end users) will comply to cybersecurity policies, practices and directives, but this trust should be earned by ensuring stakeholders are trained (cybersecurity awareness trainings, and regular certifications, and cybersecurity seminars and drills) and equipped (cybersecurity policies, good practice guides and repo are made available to all stakeholders)? The effort is harder with inter-cybersecurity practices as trust between partners, vendors and consumers are much harder to attest. Trust between organisations can be earned through demonstration of good cybersecurity practices through certifications, compliance attestation to cybersecurity and regulatory directives and mandates (e.g., ISO27001, PCI DSS, GDPR etc.,), but trust between organisation and consumers still needs to be investigated. Recent study has shown that consumers do not only care about an organisation's cybersecurity practices (how excellent they are in managing cybersecurity risk) but also the organisation's position and support for sustainability, and their position around ethics, equality, diversity and inclusion. Equality is cited as the number one consumer value (i.e., one of three top consumer trends for 2021) according to Gartner research (Omale G., 2020). In fact, these factors are becoming more important to Gen Z consumers than ever before (Duerst M. and Eichhorn E. 2020). We argue that understanding social domain dynamics to the ecosystem offer insights into the many factors that should be considered when developing a holistic cybersecurity programme that not only addresses current problems but offers foresight into the future situations. The Social Domain abridges the Cultural Domain in the sense that it also intercepts the decision making and judgement of the people, their cognitive processes and thoughts. Since people aspects are dealt with in the Cultural Domain, one might ask the relationship between the Cultural Domain (a.k.a. cognitive domain) and the Social Domain? Take for example, government digital and social structures such as e-government, e-voting etc., such important structures are reliant on digital, social and trust infrastructures, which if abused will have equal, if not overwhelming ramifications to the government as would physical or political domains. We argue that all the domains are important and do need consideration, understanding and appreciation of their risks and rewards. Their interdependence must not be overlooked, too.

The *Political Domain* deals with policy, regulation, directives, legal, legislative and corporate governance structures (see Table 1). These include local, provincial, National, lateral, multilateral and international regulation and mandates. They shape and influence how organisations and nations operate, and how we should defend society and citizens. Also, the Political Domain influences the provisos and prerogatives under which bilateral and multilateral mandates operate. The Political

454Domain, unlike all the other domains, can have an overwhelming impact on the stability, security and safety of a country and its citizens, and consequently, to its neighboring nations and continents. The Political Domain can cause ripples that propagates to the other domains, and an instability in the Political Domain can destabilize the other domains. Tension in the political sphere, for example, between two nations can cause economic instability and impact economic conditions, cultural and social strive, and certainly a reaction in the Cyber Domain. As recent studies suggest, the Cyber Domain has become the battleground for the realisation of political tensions and the exercising of power, control and dominion. For example, during the political tension between America and Russia in 2020, there were accounts of alleged cyber-attacks and infiltration between the two nations (CNN, 2020). Similarly, during the political tension between Russia and Estonia in 2007 (Ottis R. 2007). It is hard to divorce the impact of one domain to another since the domains are interdependent. It is not only the Cyber Domain that may be affected by tensions in the political sphere. The economy will be proportionately impacted, too. Stock market is usually the worst impacted. The United Kingdom's BREXIT referendum and the negotiations that followed with the European Union (EU) caused the pound sterling to drop, at a point, to its lowest in two years (BBC, 2016). It was also reported that the immediate aftermath of the BREXIT referendum caused the sterling to decline sharply in value (BBC, 2019).

The *Cyber Domain* deals with digital, cloud, information, internet and the intangible attributes that intercept the traditional physical matters of Land, Sea, Air and Space but that provides experiential outcomes which may be realised in both physical and non-physical domains (see Table 1). According to the Finnish Government Cyber Security Strategy (Finland, 2013), the Cyber Domain is an interdependent and multipurpose electronic data processing environment. And because of the dependence of all the other domains on Cyber this has made the Cyber Domain extremely important and equally, the most vulnerable. According to the DoD Multinational Experiment 7 (MNE7, 2013), "most of our activities and decisions in our physical world depend on information and access to cyber, which also continues to rapidly evolve faster in comparison to the other traditional and well-known domains such as Air, Sea, Land and Space. No wonder the growth of the Cyber Domain is seen as the biggest social and technological change of a lifetime (UK Cyber Strategy, 2021). The Cyber Domain is the most complex. According to (Bertoli G., 2018), "all systems within cyberspace are complex in their function and often have several external dependencies", hence it is not enough to protect what we consider as mission critical systems, we must also consider all the data sources and their dependability on existing and/or external communications channels, interfaces, links and interactions. These may be used as attack vectors.

**Table 1: Domains, Descriptions & Attributes**

| Domains | Descriptions | Attributes |
|---|---|---|
| **Physical** | The physical world that contains, houses and accommodates physical matter and materials. | Physical location, coordinates, physical substance that one can touch, feel and smell. E.g., Land, Sea, Space and Air, and for cyber physical systems, these are bridges, airports, water pipelines, IoT, national grid (electricity & gas), data centres. |
| **Cultural** | The cognitive or mental space, human and operator-oriented space. | Knowledge, perception, values, beliefs, attributes. All these relate to human aspects. Equally, group and collective (e.g., Team SA – i.e., situational awareness relating to group or team behaviour e.g., a group of operators, administrators, commander and their teams). |
| **Economic** | The business aspects, financial orientated space and supply-chain ecosystem. | Finance, gain, loss, impact to a nation's economic wellbeing, equity market, stock or share prices, business models and bargaining power. |
| **Social** | The social relational space, especially around nations, geographies, communities, teams or groups. | Trust, sustainability, environmental, social, relationships and structures, groups and teams' dynamics. |
| **Political** | The political, regulatory and compliance space. | Policy, regulation and governance, leadership, legal, compliance and directive imperatives. |
| **Cyber** | The abstract space of information, communication and intelligence. | Information, communication, intelligence, cyber incidents, cyber-attacks, exploits and data breaches. |

The interconnection of the Physical Domain to the Cyber Domain (e.g., through microwave and satellite communications or physically connecting them with wired and wireless communications, and/or via the sea with submarines) have created a 'centrality phenomenon' (where cyber is seen to be in the middle connecting every domain, and hence making it significantly important). For example, the Cyber Domain is now used to realise cyber harm in all the other domains, such as the physical, cultural, economic, social and political domains. It is a continuum for both 'good' and 'bad' in relative terms. We argue that strategic Cyber Foresight can only be gained, especially for a robust framework that could be used to support government, organisation and nation, when all the domains discussed in this paper are considered and impacted. The consideration of only a single domain will be a fruitless effort.

### B. Overarching Principles of the Framework

In this section, we discuss the overarching principles underpinning the framework, and the rationale for its consideration. While the proposed framework is conceptual, any realisation of the framework should consider principles that should guide its implementation (guiding principles). The following overarching principles support the proposed framework in achieving its purposes:

- *Multidimensional*: The framework should consider the different domains that exist and impact these so that a holistic understanding of cybersecurity can be achieved. Failure to consider multiple domains could lead to a framework being constrained and consequently unable to address risks arising from domains that were not considered. Furthermore, a multidimensional approach enables strategic foresight, which is gained by understanding and impacting situations from the different domains. As studies have shown, cybercrime and cyber incidents can be technically, socially, economically, legally and physically oriented (Gandhi R., et.al. 2011), therefore a one-dimensional approach is neither capable of addressing emerging and modern cyberattacks nor able to reflect geo-political, economic or socially oriented tensions that could result in the realisation of physical harm, abuse, extortion and exploitation in the Cyber Domain.
- *Integration:* The domains and components of the framework should be integrated. Integration allows the framework to function as a composite unit rather than individual and localised components that are silos.
- *Automation:* The framework should be automated because manual analysis will lag in time of the responses needed to address current and emerging cybersecurity threats.
- *Responsiveness:* The framework should be responsive or adaptable because cybersecurity programmes is not a one-size-fits-all proposition.
- *Adaptive:* A good framework is one that is adaptive so that different organisations and institutions of varying sizes and risk appetite can still employ it.
- *Coordination:* The framework should be coordinated, meaning the need for human inputs and decision making (e.g., human in the loop). Coordination of cybersecurity incident management, investigation of data breaches, decision making for incident response, geo-political escalation and sanctions and diplomatic relationships all require human coordination oversight. That said, coordination leverages automation in technology workflows and processes for speed, accuracy and efficiency.

### C. Factors

Our methodological approach using Attribute Listing is shown in Fig. 3. Through using Attribute Listing matrix and analysis methodology we derived the Factors. We then organised and categorised them into four distinct groups. The methodology allowed us to list all possible attributes and features stemming from the areas we considered. Four overarching categories are identified **- Business, Operational, Technology** and **Human** (**BOTH**) Factors, which is the resultant outcome of the approach (see Fig. 4).

It is pertinent to note that *Security*, P*rivacy* and *Assurance* (see light green horizontal of Fig. 4) neither appeared in any specific quadrant nor included in any of the Factors. We argue that security, privacy and information assurance are intrinsic features that must be considered in their own rights for all the four principal factors (BOTH Factors). For example, Operations Factors must consider information security, privacy and assurance requirements with respect to the operations that the tools undertake, and the processes should be such that they align to security standards and privacy regulations and directives. Similarly, Business, Technology and Human Factors should all abide by the same guiding principles.

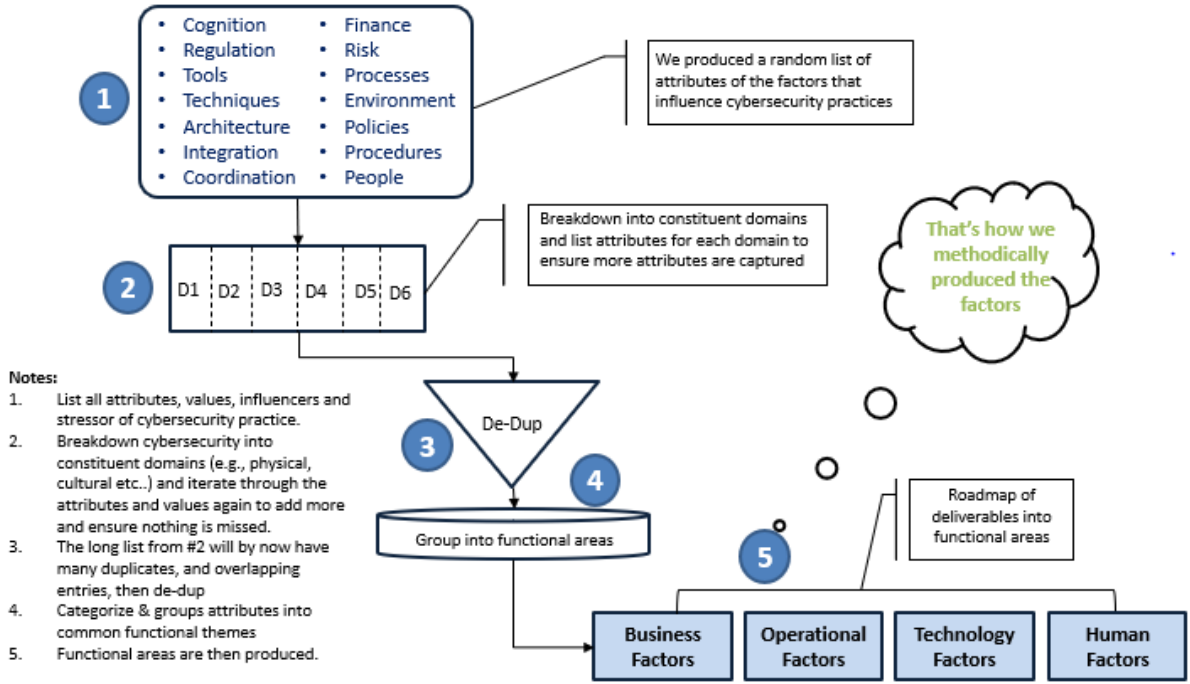

Figure 3: Methodological Approach - Attribute Listing

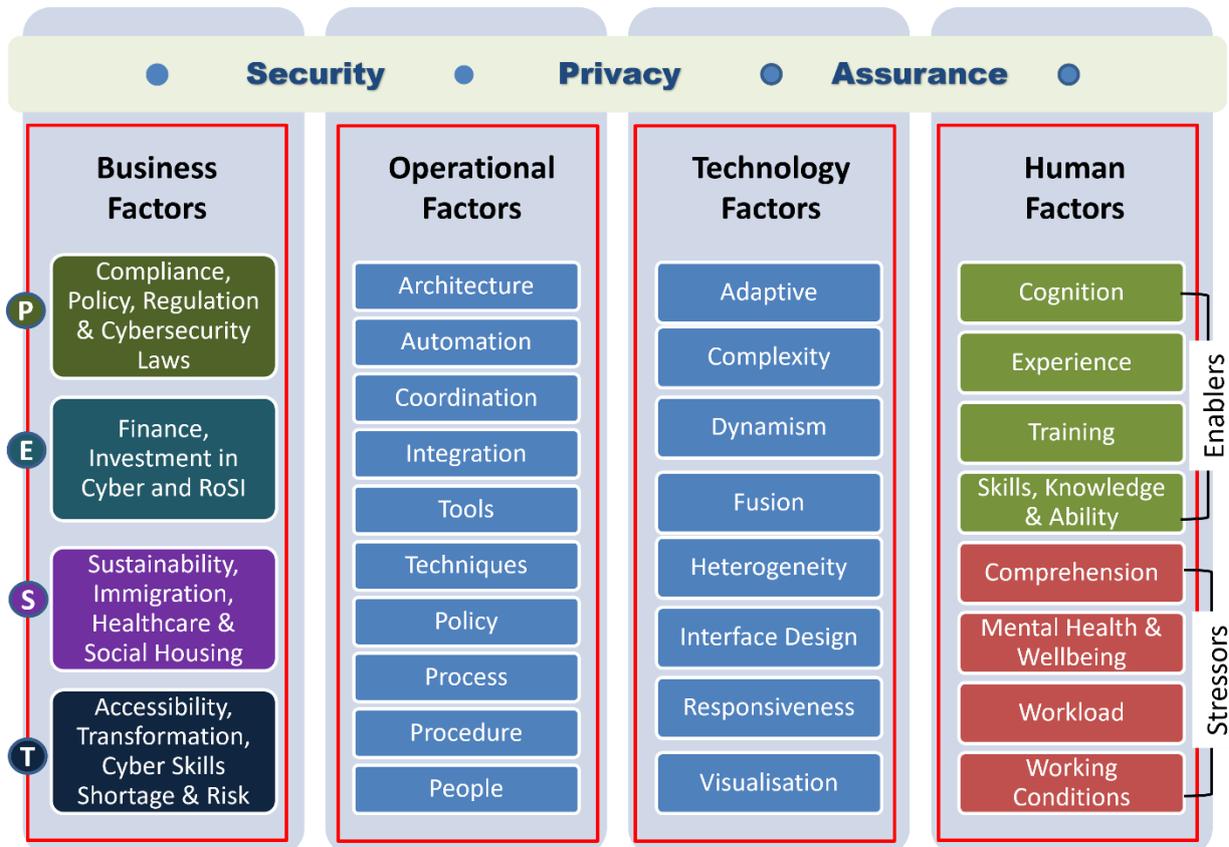

Figure 4: BOTH Factors

### C1. Business Factors for Cyber Foresight

To explore the Business Factors, we used PEST (Political, Economic, Social and Technology) analysis in discussing the overarching themes. Business environmental factors comprise both internal and external factors that influence businesses. External factors range from economic, social, legal, technological to political factors (Kotler P. and Armstrong G., 2004; Paypervids, 2016). Some studies consider *competitiveness* as an external factor, for example PESTEC - Political, Economic, Social, Technological, Environmental and Competitiveness (BBC, 2019). Internal factors include financial resources, physical resources, human resources, access to natural resources and organisational processes and innovation (Pestle Analysis, 2015). Our ability to understand current cyber situations or circumstances in order that we may be able to predict, and address future or emerging cybersecurity situations will depend on identifying the pertinent business and environmental factors that allow us to gain insight of our business environment. In this study we consider factors that influence cyber foresight from business operating environment lens. Our analysis is conducted using PEST as shown in Table 2.

**Table 2: PEST Analysis of Business Factors for Cyber Foresight**

| (P) Political | (E) Economic | (S) Social | (T) Technological |
|---|---|---|---|
| *Compliance* | *Finance* | *Sustainability* | *Accessibility and Automation* |
| Cyber disclosure compliance directive (a.k.a. Breach Disclosure) would mean that organisations located in jurisdictions and/or countries where security breach disclosures are enacted, and mandatory must report any breaches to the authorities.<br><br>While Breach Disclosure mandates are a good thing and bodes well with the principle of honesty and transparency, it could however spell doom to some companies. It is pertinent to note that the real impact of a security breach is not the total cost that may be required to protect impacted system, recover stolen data, or prevent the attack happening again or the cost of forensic investigation, but the real cost to the impacted business is often realised on the impact to their stock market, the impact to their bottom line, shares and stocks, and unfortunately, some companies have had to close down as a result. | Finance relating to monetary resources required to establish and operate organisational or national cyber programmes. The whole life cost (TCO – total cost of ownership) should include the CAPEX required to build and establish the programme, while the OPEX is the continual operating cost, which is a yearly cost, budgeted to at least five years.<br><br>For national cyber programmes, the government or sponsoring bodies (e.g., in the UK, HM Treasury) would need to review and approve funding for the business case for the establishment of such programmes. While organisational cyber programme business cases would need to be reviewed and approved by their respective organisational finance board or senior management team, or whichever body that has this authority and responsibility within their organisations. | The environmental, social and governance responsibilities is a new corporate imperative. ESG reporting is receiving scrutiny from both external and internal stakeholders alike. Employees are keen to understand the position of the company they work for on important socials issues like health and safety, diversity, equity and inclusion; environmental issues such as renewable energy, water use, climate etc., and on governance, they are interested in pay equality, board composition and racial minority diversity (Gartner, 2021). These are extremely important societal issues. | Accessibility requirements at the workplace - use of accessibility tools such as accessibility keyboards, dashboards (screens), brail, hearing loops, wheelchair access etc. could help improve affective computing, business processes and cybersecurity foresight.<br><br>Automation allows business processes and operations to be conducted swiftly leveraging technological advances and capabilities such as RPA, Machine Learning (ML) and Blockchain. Integration aids efficiency in modern business and operational technology, allowing synergy in business processes to be cooperative. This means that business processes and workflows can be automated which encourages business efficiencies. |
| *Policy* | *Investments in Cyber* | *Immigration & Border Control ('The Big Issues')* | *Transformation* |
| Government policy on profiling resulting from current incidents of covertly manning and using personal data to influence circumstances (e.g., Cambridge Analytica, Facebook) could mean that cyber | Organisations may not have enough budget for cybersecurity centres or programmes, and this may mean essential components of the cyber programmes may not be developed or started. | Immigration control stemming from political policies do often cause social issues, for example, the United Kingdom BREXIT referendum, could have a potential impact to migrants, and could worsen the existing | Technological transformations could improve business processes and operations allowing cyber foresight to be readily gained. Business transformation in form of digitalisation allows many business |
| programmes would be unable to utilise certain data analytics tool or use relevant data that could enrich intelligence or cyber foresight. In the case of the Mission, agencies may therefore need to resort to the invocation of national security mandates in order that they may be allowed to process high volumes of sensitive personal data that could have ordinarily been constrained by the data privacy mandates (e.g., DPA and GDPR requirements). | Cost of developing national cyber programmes is likely to rise due to poor national economic outlook, and the cost of attracting skilled cybersecurity SMEs is likely to rise due to cyber skills shortage. Furthermore, the ability to sustain cyber programmes or national fusion centre is a challenge since these programmes are not for profit and therefore are operated on a different paradigm to how industry projects or programmes are run. Longevity of national cybersecurity programmes are dependent on government sponsorships and funding. Should funding not be secured some government programmes do not continue. And with cuts in national budget (reduction in funding) to central government departments this would impact cybersecurity programmes. | shortage in cyber skills. This is because skilled workers may immigrate to other countries other than the UK leading to widening skills shortage gap in the UK.<br><br>The US/Mexican border immigration control is another such example, and may result in many unintended consequences, and one of such, is that skilled workers may leave the US and equally, the US may find it challenging to attract fresh skilled immigrants. | processes and operations to be digitalised, automated and integrated.<br><br>Digitalisation allows business processes to be conducted from anywhere and everywhere, which means no geographic constraints or time zone issues, this effectively means increased effectiveness, and/or productivity. And innovation could drive efficient and optimised business processes through automation, integration and intelligence. |
| Regulation & Cybersecurity Laws | *Return on Security Investment (RoSI)* | *Healthcare and Social Housing* | *Cyber Skills Shortage and Risk* |
| Regulations like GDPR means that user consent must be required before data collection, and this may impact data collection and analysis regimes, especially around cybersecurity monitoring.<br><br>National cybersecurity laws provide sovereign and safety net to countries to protect their intellectual property (IP), critical national infrastructures (CNI), government and infringements and interference from nation states. National cybersecurity laws are powerful and while it is a force for good, it has also been used in authoritarian nationals to curb, and control private organisations (Chinese Cybersecurity Law). The recent crackdown of Tech and Ed-Tech publicly listed companies, e.g., DiDi, Tencent, Alibaba etc. by China on cybersecurity concerns impacted the market, and demonstrated how these laws can impact society in general (Cheng E. (2021)) | Return on security investment can be measured through technological advances, allowing business operations, activities and risk to be appropriately assessed and measured. It is easy to measure certain things, especially tangible things, for example, profit and loss of consumables and goods; however, it is challenging to measure intangible things, for example services as they are subjective. It is even harder when it comes to measuring cybersecurity profitability. Because of this, we believe RoSI should be assessed based on cyber KPIs where certain key performance indicators are used to measure RoSI. Cyber foresight cannot be measured by a single parameter, it is a culmination of several useful KPIs. | The cost of healthcare in certain countries makes it unaffordable for most, especially those living in cities where healthcare is either not free or subsidized, for fresh graduates whose wages are not enough and hence unable to afford it. This is likely to have negative impact to organisations operating in such cities and the ability to attract and retain skilled workers.<br><br>The lack of affordable housing is likely to impact the young working force, and may impact their choices of living or location, as most people will look for employment where they can afford, and this may invariably bias employment. For example, In the UK, while London may be attractive to young people, unfortunately due to the high cost of living exacerbated by unaffordable housing is leading people to move out of London, and this is having social impact to the city and equally to industries in the city as they must pay higher wages than is paid to colleagues in other parts of the country causing also social divide. | Cyber skills shortage may mean that cyber programmes may struggle to attract the right cyber analysts or cyber personnel to work at the centres.<br><br>Through technology, risk assessment and risk management can be applied to ensure business operations are better understood. |

### C2. Operational Factors for Cyber Foresight

Operational Factors deal with operational aspects of the cybersecurity programme. Operational Factors somewhat realise the contributions of the Technology Factors when implemented correctly and ensure Business Factors are met. Operational Factors range from Operational Technology (OT), tools to techniques employed by operations and change management, as shown in Fig. 4. The Operational Factors discussed in this paper are derived using *Attribute Listing* and F*iltration. Attribute Listing* allowed us to enumerate all operational factors, attributes and features, and by filtering against duplications, we removed factors that are either similar, duplicates or overlapped with existing features covered in other parts of the framework (see Fig. 3). The resultant Operational Factors are discussed as follows:

- **Architecture**

Operational architectures are 'live' architectures that are in the production environment of the organisation and are used to process and execute live services and systems. They include patterns, designs, interfaces and building blocks (i.e., architecture artefacts) that underpin business and IT technologies. Operational architectures for systems and services are underpinned on the business requirements of the organisation, and to support and ensure that business needs are achieved. Operational architectures should use baselined architecture patterns, architecture designs and architecture blueprints that have been assessed, approved, tested and baselined; and are in use in parts of the organisation. We gain operational efficiencies and time savings by reusing existing architecture artefacts, and by ensuring continuous assurance of architecture contents. This means architecture repository is regularly updated with new and emerging secure by design architecture collaterals and artefacts. Operational architectures should describe tasks that they accomplish, operational elements and information flows and patterns that should be used for (Dictionary of Military and Associated Terms, 2005), and they should be continuously reviewed and managed through architecture change boards, e.g., DevOps (Development and Operations) and CI/CD (Continuous Integration and Continuous Development) pipelines. This is so that changes in the architecture space have full audit trail, authorisation and accountability.

- **Automation**

According to (Onwubiko C. and Owens J.T., 2011), the advent of powerful computing systems and the possibility of real-time processing have made automation of technologically enabled and situation-aware tasks much more attractive and accessible. The processing of data and information using graphics processing units (GPUs) in addition to its affordability have made processing of big data a reality. Nowadays there is a real desire to automate almost everything ranging from physical, logical, perceptual and even cognitive tasks. This is invariably driven by the needs of modern society, the quest to process things at ultra-fast speed as our lives move at a tremendous pace, and technology is used to power and automate processes, e.g., robotic process automation is now used to automate the ordinary everyday tasks, such as check email, reply to emails, tweet, post Facebook updates, write blog posts and aggregate news feeds. Automation brings far greater advantages in cyber, especially around speed, accuracy and precision. We need these attributes if we are to gain Cyber Foresight. Take for example, in information operation in warfare, being able to analyse huge volumes of data at a greater pace, accuracy and precision can become the unique selling point (USP) that differentiates Cyber Foresight over one mission and their opponents. Automation allows for tasks, processes and routines to be mechanised, operationalized and industrialised to gain greater depth and efficiency. In today's world of information overload, big data, cloud computing, and the plethora of social media all competing for both our time and attention, and huge volumes of data to process, it would be challenging to gain enhanced situational awareness of the threat landscape let alone gain Cyber Foresight across the interwoven domains without automation.

- **Coordination**

Coordination relates to the human aspect that deals with the organisation of the different and disparate elements of a complex activity to enable them to work together coherently and effectively (Dictionary Online, 2019). Coordination is an intrinsic human cognitive function through which elements of complex tasks, operations and maneuver are arranged, organised, fused and/or managed collectively and collaboratively to achieve a common and desired goal. Coordination is applied to both cyber and non-cyber related activities. But since this paper is focused on Cyber Foresight, our examples and explanations are tailored to Cyber, where coordination is an extremely important feature. Take for example, cyber operators and analysts coordinate cyber incident response and crisis management. These tasks on their own are complex, time and cognitively demanding. Each aspect of Cyber is interdependent, and it is this interdependence that necessitates coordination. For instance, responding to cyber incidents requires a coordination of various tasks, e.g., security monitoring, correlation of security signals, and cyber incident response. Cyber incident responders perform activities such as gathering, preparation of digital evidence and the preservation of digital evidence, with the incident management helpdesk, and the senior management team who make decisions on the approach and possible cause of action and including authorisation for reporting of the breach to national authorities, where applicable. While cyber operators may use technology and automation to perform many of the complex tasks, allowing and leveraging machine intelligence, speed and accuracy, coordination is, and will remain largely a human factor attribution. We argue that very limited foresight is possible in a chaotic, uncoordinated and siloed ecosystem. We believe foresight will be enhanced in a cooperative endeavor based on human to system (H→S) relationship. A reliance on a singular aspect, say machine alone (without humans), will not provide the required levels of foresight. A coordination across the domains allows insights and situational awareness to be gained through monitoring and intelligence sharing of the ecosystem.

- **Integration**

To achieve foresight of any kind, cyber or otherwise, systems integration and automation are key. Operational technologies and systems ought to be integrated so that they form a cooperative and co-existing system of systems that delivers the overarching functionalities, interaction and business processing. With systems integration, disparate and diverse systems, components and subsystems that would have ordinarily existed as separate, isolated and siloed systems creating fragmentation are brought together in a way that allowed services operated across them to be cooperative and coordinated. Cyber defences form a layer of protection (defense in depth) only when they are integrated. Integration can be achieved in hardware, software, programming and hybrid, for example, systems integration of CPU to motherboard, keyboard and monitor is achieved through hardware integration, network integration can be hardware or software, process integration can be achieved through application programming interfaces (API), while robotic process automation can be achieved in hardware, software and hybrid (e.g., cyber physical systems). Integration and automation of cyber defence systems is an absolute business requirement and considering the plethora of systems and applications that the CSOC monitors, it will be challenging, if not impossible, to monitor such myriad of systems and networks without systems integration and process automation of the CSOC central collection infrastructure. Besides, systems integration helps offer service efficient and customer value-add through improved product quality and performance (Vonderembse M. A. et al, 1997).

- **Tools**

These are the technical and technology capabilities deployed in the framework that allow operational aspects of the service to be swiftly executed and processed. Without capable tools in place, it will be challenging to realise operational efficiencies or meet some of the overarching business goals. For example, using the case of a security operations centre programme, as a use case, without appropriate technical tools, SOC operations will lag in time and consequently fail to achieve its business goal of real-time continuous monitoring. This is because it will be overwhelmed with high volumes of logs that it could not manually process without the use of technical correlation and analysis tools such as security event and information management (SIEM) tool. Tools aid technical effectiveness of the programme and help in processing data and events continuously allowing the programme to be efficient and to gain enhanced situational awareness of the ecosystem. Tools selection and choice must be depending on achieving the features articulated by the Technology Factors (see Section C3), some of those include, e.g., Tools must be situation-aware, have the prerequisite interfaces, be automated allowing for orchestration and workflow processes. It should be integrated and support multiple interface types such as API, native etc., and should be smart and intelligent. It is important that operational tools have the capability to process huge amounts of data, easy to use, and portable. These are some of the features that guide tools selection and choice.

- **Techniques**

This relates to the approaches used to operationally analyse, process and conduct activities. Such techniques might be approaches to SOC operations, analyst work, cyber incident event processing or incident management coordination. We argue that techniques that allow cyber programmes to leverage efficiencies in automation, workflow and orchestration offer the SOC the much-needed pace in keeping up with cyber incident response which is very much dependent on speed and accuracy. Industry best security and technology practices in this space include the MITRE ATT&CK framework (Mitre, 2017), the Lockheed Martin Cyber Attack Kill-chain (Lockheed Martin, 2016) etc. Operational techniques should be driven by operational efficiency, performance, speed, accuracy and precision.

- **Policy, Process and Procedure**

Policy, process and procedure are the foundation for operational technology without which it will be infeasible to conduct operational tasks. Policy provides guidelines, procedure stipulates the low level, 'how' of applying the policy, while the process helps industrialise the procedure and therefore allows a consistent approach to be followed. Like other operational tasks, Cyber operation is no different. Take the CSOC for example, they need policies, processes and procedures to operate the SOC service. These may include policies on a wide range of tasks, from simple to complex tasks, such as Joiners, Movers and Leavers (JML) policy, Access provision and Deprovision policy, User Rights Management policy etc. They need procedures to follow, for example, cyber incident response procedure, major incident management procedure, etc., and likewise, they need to have processes in place that allow consistency across various repetitive tasks, at least, for example, cyber incident playbook process, access requisition process, account creation process etc. Cyber programmes, especially CSOC should have operational policies, processes and procedures that they follow in a systematic and consistent manner that enable service efficiencies. To stay ahead of the game, we argue that cyber programmes should have very robust policies, procedures and processes and these need to be relevant, current and maintained, and most importantly, made readily available to staff. Often, processes may exist, but staff are not aware of them due to poor communication or limited document management access or both. It is pertinent to note that the relevance of these artefacts depends on several factors, namely:

a) All staff should be made aware that policies exist
b) All staff should have access to policy documents
c) All staff should be trained on the use of policy artefacts
d) All staff must be briefed of the consequences abuse of policy.

- **People**

Operations people comprise staff who are responsible and accountable for maintaining the operational day to day activities of the organisation ensuring that operational systems are effective and perform in accordance to stipulated functional and business requirements and supports the organisations services. There are many categories of operational staff, ranging from cyber operators, operational staff, cyber incident responders, administrators, analysts to management teams. These people are incredibly important because they deal with the overall operational, administrative and change management aspects of the service operations and maintenance. They perform the business-as-usual tasks, operate the technologies that drive the processes and ensure that the tools and technologies are maintained, operated and serviced. For cybersecurity tools, the situation is even much more critical and impacting. For example, if cyber defence systems are not continuously updated or patched, vulnerabilities may exist and this could result in the safeguards being exploited and further, they may then be used to compromise the wider ecosystem. Human operators conduct cyber incident and crisis management, monitor the infrastructures, networks and systems, carry out analysis and investigations when a security breach occurs, and take part in decision making, escalations and reporting. People will always be required in most manner of endeavors in some form or another. This is most pertinent with cyber; while there exist machine learning models that can predict cyber-attacks, artificial intelligence models that can recognise speeches and deep learning models that can investigate, recommend and optimise choices for humans, there are still, at least for now, areas and use cases where human operators are needed and may still be needed, for example, in decision making, escalations, cybersecurity investigations, and prosecutions etc. There is no doubt that human operators will have to depend on cyber- physical systems, machine learning models etc. for swifter, more precise and optimised choices, however, it will be a case of interdependency than replacement or displacement. We see a cooperative situation where humans leverage technological power, prowess, speed and accuracy in human decision making, prioritisation and resolution.

### C3.     Technology Factors for Cyber Foresight

Technological Factors are technical requirements and capabilities that should be considered to gain Cyber Foresight for organisational or National cyber programmes. These factors are characteristic of the necessary *functional* and *nonfunctional requirements*. While we have not attempted to distinguish between functional and nonfunctional attributes, we have discussed both knowing that regardless, both are required in achieving and maintaining Cyber Foresight for the organisation. Factors we have considered for the technological aspects for achieving Cyber Foresight are obtained by applying Attribute Listing – a methodology that allows for features of objects to be listed, including low level attributes, so that by combining these attributes new concepts may emerge or be identified. These attributes include:

- **Adaptive**

It is without a doubt that we are now beyond the 'valley of death' in AI and ML. *Valley of death* is a business idiomatic description to highlight technological maturity and investment take-off, the 'death' occurs when technological advancements, for some reason, do not take-off or lead to a successful business beyond proof of concepts and laboratory demonstrations. AI, which has been in existence for several years has been in the 'valley of death' until recently. It has since recently led to successful and sustainable businesses. AI and ML (and Deep Learning) have been applied to many problem spaces to revolutionise operations. In these cases, their implementation, including the feature engineering and the algorithms must be adaptive as each problem space is unique and the use cases are varied and different. While it is not new to deploy automation to achieve faster and better efficiency, with data and even, big data being readily available now compared to several years ago, the problem space has shifted from one that needs data to automate tasks so that intelligence and efficiencies can be gained to one that now have data, and therefore must adapt automation in the right possible ways to leverage efficiencies and effectiveness. To gain enhanced situational awareness and Cyber Foresight, we argue that our technologies or the technologies we deploy in our cyber programmes should be adaptive. This includes the tools, techniques, and workflows. It is only through adaptation can we exploit the opportunities and take advantages that big data and the revolutionized AI offer us. We leverage these tools to gain insights and cyber superiority over our adversaries.

- **Complexity**

We live at a time of unprecedented complex ecosystem. The technologies are complex, data is complex, operational processes involve far more stakeholders than ever before, with far reaching supply-chain ecosystem covering geographies and economies that were not physically and otherwise integrated prior. Entities now involve both humans and bots, and the activities and interactions of both are difficult to differentiate. Geographic reach is no longer a constraint, there are various technologies overlaid on the Internet that have addressed these issues. For example, realtime communications technologies, geographic visualization tools, and geo-location technologies that can track entities to their nearest post codes, street-level location etc. at high precision and dimension now exist. All of these has made gaining insights and understanding of our environment extremely pertinent and equally challenging. Complexity is not a singular or localised issue. It is seen in every aspect of society, in Economic, Social, Cultural, Political and Physical domains. Cognition like cyber is very complex and convoluted. Mental processes are challenging to understand or program, likewise, all the other domains comprise of complex interlaying of knowledge and activities that are intriguing to comprehend. The fact that these domains are interdependent and interwoven, as no domain is isolated or an island on its own, has exacerbated the complexity of inter-domain interaction and multi-domain relationships. Gaining Cyber Foresight or maintaining enhanced situational awareness across these domains is challenging.

- **Dynamism**

Each of the domains namely Physical, Social, Cultural, Economic, Political and Cyber are different and so are their requirements, properties and attributes. For example, the Physical domains of Land, Sea, Air and Space may not be as dynamic as say the Cyber domain. The Cyber domain processes huge volumes of data, information and intelligence using computer networks and communications systems, and their operations are dynamic, complex and often abrupt as shown by the varying levels of events or packets they process, transmit and store. Communications are swift, realtime, dynamic and continuous, and computer networks process high volumes of data, information and events, and these can often be unpredictable, depending on the activities and user base they operate or transactions occurring daily. Modern computer systems and networks are built with extremely fast processing capabilities, such as GPU to boost the performance of video and graphics. They comprise multi-core CPUs, large memory banks and disk facilities that enable turbo fast processing of data and information. No wonder why some networking devices, such as routers, switches and firewalls can process thousands of events per second (EPS), analyse terabyte and even petabytes of data in minutes. Analysis of any sort can now happen in seconds or tenths of minutes instead of hours. The mean time to detect incidents are continuously reduced and lowered; databases are now graph-based instead of the traditional relational-based architecture, allowing for faster input and output operations. The advent of cloud computing has taken this to a new and unprecedented levels as the power of parallel processing and infinite compute augmentation can be leveraged to render data and information at speeds never seen before.

- **Fusion**

Fusion is a technology requirement for Cyber Foresight. Data fusion is a technique to aggregate disparate sets of evidence regarding a perceived situation to better understand its relationships. It is a capability that allows correlation and cross-correlation of multiple sensors or data sources across the different domains, so that information from one domain can be related with information across the other domains. Multi-data fusion allows for enhanced situational awareness to be gained by allowing information from one domain to be cross correlated among the other domains, allowing for greater visibility, understanding and comprehension of the relationships of a single incident across the domains. For example, by fusing information across the domains, and by correlation, incidents that would have gone undetected or incidents that would seem unrelated can now be identified and their relationship substantiated, and the wider impact can equally be uncovered and understood. Fusion allows for data from multiple heterogeneous sources to be combined to obtain better and higher degrees of accuracy and richer inferences than those obtained from a single source. According to (Hall D. L., and McMullen S. A. H., 2004) fusion is a collection of framework, theory, tools and techniques for exploiting the synergy in the data, information or evidence acquired from multiple sources, such as sensors, databases, networks and humans that helps us better understand a phenomenon and enhance intelligence. We argue that data fusion is better in detecting wide-spreading and enterprise-wide situations (network faults, threats and attacks) targeting networks or enterprises as the understanding or intelligence gained from the analysis of multiple data sources outperform that which is gleaned from a single source. This is a fundamental preposition for Cyber Foresight, as enhanced situational awareness is far more plausible with fusion across the domains than analysis of single items against a single domain.

- **Heterogeneity**

Heterogeneity is a fact of all facets of life. Different tools, mechanisms and processes are used, and no single organisation can claim to use the same tools (homogenous) or same processes for everything. Certainly, in the Cyber domain, we leverage the power, quality and veracity of heterogeneity to obtain the best of breed in cybersecurity monitoring. Communications and computer networks, infrastructure, operating systems and applications are monitored using heterogeneous tools and techniques to analyse traffic to detect vulnerabilities, cyber-attacks and data breaches. Heterogeneous data sources observe, process and detect dynamic changes in the network. E.g., the use of intrusion detection systems (IDS), firewalls, anti-virus systems and boundary security controls to collate security information events observed in the network, and the use of security information and event management (SIEM) tools to analyse, correlate and detect cyber incidents. Heterogeneity is applicable to the other domains, such as Economic, Cultural, Social and Physical domains. Through the use heterogeneous controls enhanced situational awareness and Cyber Foresight can be gained. Since no single control can identify all security threats, therefore, it is recommended to use heterogeneous controls to detect a wide range of threats perceived in the network, which is not possible from a single control perspective. We argue that observation and intelligence from myriad of multiple heterogeneous sources offer richer Cyber Foresight picture than that from a single source. Similarly, enhanced situational awareness is gained by integrating, correlating and cross-correlating information from the six domains than that gained from a single domain. This is the premier of our multi-domain and multi-dimensional approach to cyber situational awareness for Cyber Foresight.

- **Interface Design (a.k.a. User Interface)**

With the huge volumes of information being generated, processed and outputted by computer network systems to the users/operators for decision making, then interface designs have become as equally important as the complex processing capabilities inbuilt into such systems. Human operators should be able to use the system to make inferences, and the interface designs as much as the output should be intuitive and easy to understand (usability and enhanced user experience). These systems use a plethora of technologies and software to implement complex business logic, some of which involve a finite number of background processes that are transparent to the operator, and therefore make it challenging for the operator to swiftly identify sequence of erroneous actions or detect a fault in the system or spot when an abnormal situation is happening. To assist operators to detect, diagnose and remedy abnormal situations swiftly, complex computer systems should provide interfaces that enable human computer interaction. Operator situational awareness will be enhanced if systems can provide interactive UI that enable human interaction. These UIs enable Human Computer Interface (HCI) and feedback loop. Whatever the UI, systems and their components need to interact (intra and inter

65

communications), integrate and interoperate either as autonomous or interdependent systems. And this is one of the reasons why UI design has become a significant factor when assessing situational awareness in systems. According to Endsley M., and Garland D. J., (2000), one of the primary reasons for measuring SA has been for the purpose of evaluating new system and interface designs. HCI designs do affect operator performance and system safety. HCI design is shown to have a profound effect on safety assurance, particularly during emergency situations (Sandom C., 1999). For example, when emergencies arise and system operators must react swiftly and accurately, the situational awareness of the operator is critical to their ability to make decisions, revise plans and to act purposefully to correct the abnormal situation. This sentiment emphasises the importance of designing UIs to support situational awareness, especially in complex and dynamically changing environment such as in security monitoring. The quality of a system's UI (e.g., aesthetics and accessibility) determines the degree of human interaction possible, such as touch, responsiveness, aural etc.

- **Responsiveness**

Responsiveness is the ability to present data and information, especially graphics and graphical-based imagery, meaningfully across multiple platforms without the need to manually reformat the layout or appearance. Responsiveness is a very powerful feature of Mobile and Tablets that allows these portable devices to present information in readable and graphical layouts without distortion, especially for graphical user interfaces across multiple screen sizes and dimensions. This is phenomenal when considering that most of the information and graphics that are now displayed nicely on Mobile and Tablets would have been previously challenging to display without manual adjustments of the images to predefined sizes and shapes. This feature on its own is important, but considering the number of Mobile users to date, which is, 4.58 billion (Statista, 2019), and this number is growing, makes this feature extreme pertinent. Mobile and Tablets are now mainstream channels of information consumption, and in fact, more people access information and services on Mobile and Tablets than on desktops. Cyber Foresight and situational awareness of the environment and domain should therefore consider access from all forms of channels. Mobile and Tablets are used to disseminate and receive communications of all forms and classifications of information. Their ability to render services, especially satellite imagery, Geographic Information Systems (GIS) and graphical objects undistorted is an important feature to consider. We argue that the ubiquity of Mobile and Tablets platforms, and their uses in modern communications make them extremely pertinent for technological considerations to gain enhanced situational awareness and Cyber Foresight. This also implies to technological impact across all platforms – IoT, Radar, Sensors, Desktop, Mobile and Tablets.

- **Visualisation**

Visualisation is a powerful tool to understanding relationships, not just of one domain but across multiple domains. For example, as illustrated in Fig. 6 & Fig. 7, these figures help visualise intradomain and interdomain relationships, providing better understanding than textual because relationships can be seen and visualised. We use graphical visualisation to understand relationships in situations, connections, patterns and outliers. Security visualisation has been used to visualise patterns in cyber-attack information, threat intelligence content (e.g., attack graph) to assist security analysts to swiftly spot or detect a cyber incident. Visualisation is the display of organised data and information into meaningful patterns or sequence of actions and activities that are visualised on a dashboard, screen or other displays. It is shown to aid faster and better comprehension of situational awareness as visuals create lasting images of the situation and show both temporal and spatial relationships among the objects. Visualisation can be applied to any of the domains to understand patterns and relationships to quickly gain Cyber Foresight. In the Economic domain, visualisation can be applied to better understand the stock market, macroeconomics and monetary policy impacts on commodities or shares, for example. In the Political, Cultural or Social domains, it can be applied to understand social networks, human and cognitive behaviour and actions, settlement patterns and demographics. Visualisation allows network information, indicators of compromise (IoC) to be monitored and visualised providing cybersecurity analysts a rich display of ongoing or historic trend of attacks so that they can better identify relationships, associations and traffic patterns and behaviours, and to process large amounts of information concisely. According to (D'Amico A., and Kocka M., 2005) security visualisation has proven to be a valuable tool for working efficiently with complex data and maintaining situational awareness in dynamic and demanding operational environments. Visualising network activities can be useful to both decision makers and security analysts in identifying patterns of attacks, and in decision making, control selection and cause of action. For example, visual analytics is essential to obtaining enhanced situational awareness in networks (Gregoire M., and Beaudoin L., 2005), understanding endpoint-level netflow traffic in networks (Lakkaraju K. et al., 2004), and continuous monitoring (Yin X. et al., 2008).

### C4. Human Factors for Cyber Foresight

Situational awareness has always been needed for people to perform tasks effectively (Endsley M. R., and Garland D. J., 2000) regardless of the domain of application. While technology, tools and robotic process automation in recent years have been used to automate tasks, and even accomplish the tasks faster than humans and with a higher degree of accuracy, however, humans (namely, operators, analysts, and decision makers) are still needed in the ecosystem, at least for now, and hopefully, for the foreseeable future. Across the domains, anything and everything can be automated to a degree. AI and ML can now be used to 'learn' tasks that humans used to do, and even perform those tasks with improvements, however, contextual intelligence without human in the loop is still uncommon. The argument is not that it could not or would not happen, the argument is that you will still need humans to provide *citizen feature engineering* and *domain knowledge expertise* required to codify and 'learn' what good looks like or not, which is offered by a *citizen data scientist*. Even with unsupervised learning, where 'learning' is claimed to be unsupervised, it does require human input and guidance, at least of the data and the business domain of the problem to be solved. In cybersecurity, where automation and ML are now at the forefront of defence controls, especially around endpoint protection and boundary control analysis, malware detection and behavioral analytics, yet we still rely on humans for decision making, cyber incident response, incident coordination and operational security monitoring. Therefore, it is pertinent to discuss human factors regardless. We argue that human operators gain enhanced situational awareness of the domains, the environment and the context in order that Cyber Foresight can be gained through many enablers, and the impact therefore of the stressors, too. In this paper, we divide Human Factors into two categories, namely *Human Factors Enablers*, such as knowledge, skills and abilities, experience, training, cognition and group or team experience and support. The other Human Factors are what we term *Human Factor Blockers*. We explain Human Factor Blockers as factors that inhibit (or worsen) situational awareness of an individual or group of persons (e.g., shared stressors). These factors include stress, mental health and wellbeing, anxiety, workload, working conditions, health & safety and diversity, equality and inclusion (see detailed discussions on these in Section – *C4.1 and C4.2*).

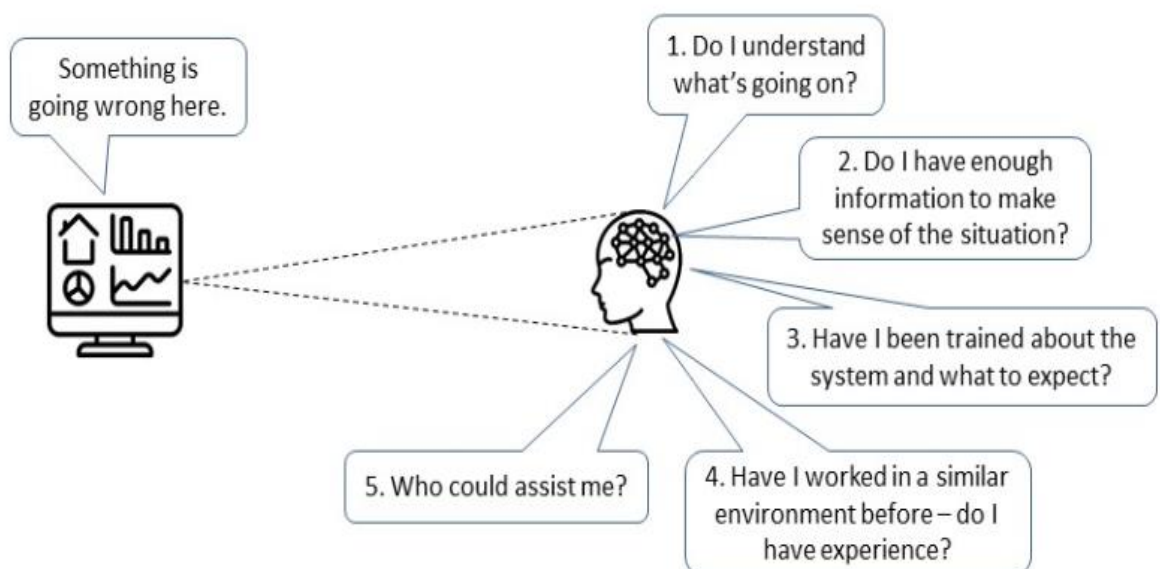

**Figure 5: Human Factors Requirements for Cyber Foresight**

Fig. 5 is our illustration of Human Factor requirements for Cyber Foresight. This picture depicts a human operator being asked to 'make sense' of dynamic and spontaneous flowing network traffic information and alert scenarios, that is, to monitor the network. First, network traffic is processed and analysed using a plethora of very complex technologies and tools, such as firewalls, antivirus systems, intrusion detection systems (IDS), flow analyser, security information and event management (SIEM) systems, etc. Each processing huge volumes of data using complex automated processes and workflows and producing information in dynamic changing but abrupt frequencies. Cues and prompts are rendered as alerts to the human operator (e.g., security analyst, SOC operator). Graphs and trends of varying colours are displayed on the dashboard (and in most cases on multiple dashboards and screens), which the analyst must check and investigate.

As shown in Fig. 5, there are several important questions the analyst needs answers. For example, given all these myriads of information and cues:
1. When do I know something has gone wrong?
2. What cues and prompts reliably show/depict a compromise or security breach? (Due to the high false positives with most IDS, this is not such a trivial question)?
3. What happens if the information or evidence produced by two or multiple sources are in conflict and non-complementary?
4. What happens if the information is incomplete or inconclusive?

In response to the above questions, there are likely to be several assumptions.
- **Assumption #1:** The analyst will be prompted automatically by the alerts or alarms that are triggered by the sources and controls, such as IDS, SIEM whenever an event/incident occurs.
- **Assumption #2:** The display (e.g., dashboard, or visualisation) will provide varying degrees of visual clues e.g., fluctuations in peaks and troughs of the trendline, changing colours of the metrics being monitored etc. to indicate abnormal changes in the traffic being monitored. There is likely to be changes in the traffic volumes, too, especially from baseline profile.
- **Assumption #3:** The helpdesk (e.g., assuming the organisation has a helpdesk function) and impacted customers might contact the analyst on phone, SMS, email or use other channels, e.g., slack, to enquire of the failure in service, network outage or incident.

The analyst will have to contend with the high rate of change, velocity, volume and veracity of the monitoring environment, and including other things beyond his control. For example, cyber-attacks happen unannounced, and no one knows when a cyber incident will occur. Also, multiple and multi-stage attacks may occur simultaneously. Furthermore, because cyber-attacks require immediate and instant response to mitigate the attack to reduce and minimise the impact, and most importantly to preserve forensic evidence which may be analysed at a later stage to understand who conducted the attack, where from, what they did, and probably if they left a backdoor to carry out future attacks or security breaches. With all of these, the analyst is faced with further questions such as:
a) Do I understand what is going on? - It is probably not going to be a binary answer of 'Yes' or 'No'. This answer will depend on several other factors, for example, internal and external factors to the user. One of the internal factors will be. Does the user possess the experience to understand what is going? Has the user been trained? A couple of the external factors to the user, will be, are there sufficient information or intelligence provided in order that the user can make sense of the situation? The information outputted by the monitoring tools is it reliable? There are many other factors that will need to be addressed to appropriately answer these questions.
b) Do I have enough information and data to make sense of the situation?
c) Is the information presented to me in a manner that is readily easy to help me quickly identify the problem?
d) Have I been appropriately trained about the monitoring system and what to expect in such an environment?
e) Have I worked in a similar environment before (e.g., prior knowledge)?
f) Who could assist me if I do not know, and to whom could I escalate to?

The above questions could be amplified in certain environments and domains but providing appropriate responses would require the continuous extraction of information and meaning about a dynamic system and its environment. The ability to combine this information with previously acquired knowledge to form a coherent mental picture, and the use of that picture in directing further perception of, anticipation of, and attention to future events (Wickens C. D., 2008). It is this process that allows Cyber Foresight to be gained of the situation, its environment and the context.

### C4.1. Human Factor Enablers
Human Factor Enablers help answer the questions posed by Fig. 5. We discuss them as follows:

- **Cognition**

We discuss cognition in relation to operators' mental ability to recognise, understand and comprehend cyber situations, e.g., an outbreak of malware, attempted intrusion, information manipulation, gamification of the monitored systems, cyber espionage, incident investigation etc. Achieving Cyber

Foresight against these cyber situations requires a combination of tech and human cognitive abilities (Endsley M. R., and Connors E., 2012). Operator cognition is as important as technology because operators' cognition is helpful in identifying and recognising situations. Cyber security experts rely on their experience, abilities and knowledge of the domain (e.g., domain experts) to understand cyber situations. Operators' cognition of situations is built upon their prior knowledge of existing situations or similar situations to the existing one. Their understanding is much better through their mental processing if they have seen, observed, handled or dealt with similar situations in the past. For example, related to observations they had seen in the past, patterns they recognised and relationships that they are familiar with. All these draws on the mental apriori of the operator. We argue that operators possess improved cognition when they are trained and experienced; obviously when they are aided through the power, pace and accuracy offered by technology the outcomes are far much better. Experienced operators leverage historical or past knowledge (mental apriori) of the situation to make better and informed decision. That is, if it is a situation the operator had previously handled, the operator is most likely to understand it better and hence be quicker and appropriate in responding to the situation. That said, the operator must first understand the situation, the nature of the attacks, what the attacks are targeting to exploit (e.g., vulnerability that may exist in the asset), and with all these able to assess the impact and possible remedies, and on how best to respond swiftly to reduce the impact. There are occasional cases where operator judgement without prior knowledge or experience of the situation may be appropriate, however, this does not negate the necessity of prior knowledge and experience.

- **Experience**

Experience is a measure of understanding a person has or possesses of a specific area or field. It does not and should not correlate to the number of years that person has been working in that field. While it is expected that the longer one works in a certain area or on some certain tasks then the better, they become at it and hence reflects their experience. This does not negate the possibility that certain individuals learn faster and perform better. In relation to Cyber Foresight, experience is key to gaining situational awareness of adversarial techniques, tactics and approach. To have cyber superiority of any kind, the mission must have experienced people who are not only able to understand the adversary but most importantly their approach, techniques and tactics, which would allow them to intercept, prevent and detect the adversaries' attempts and intrusions, and ultimately foil their intent. The adversary may target a domain or multiple domains therefore, it is important to gain insights of the weaknesses and vulnerabilities that may exist across the domains (cross domain situational assessment).

- **Training**

Training is essential in all walks of life. It is invaluable in today's technologically fast-paced modern world. The pace of technology and technological advances necessitates the need for lifelong learning and upskilling. To ensure analysts are good on their job, they need training (upskilled and reskilled). The fast pace of the modern cyber world means that new techniques, processes, methods and applications are frequently introduced. Hence operators are now required to be trained and re-trained much more frequently to cope with the increasing speed of technological advancements. Take for example, technology refresh cycle is now between 12 - 18 months (used to be 24 months or more), meaning new technologies are introduced ever so frequently. Operators of the new technologies will need to be trained on the new technologies, and while this is going on, it is likely that their ability to understand and comprehend information produced by these techs will improve over time. That said, training is not a panacea as these operators work in dynamically changing environments, therefore, they are expected not only to know the systems and their outputs, but also, be experts in interpreting the meaning of the outcomes from these systems. Training Needs Analysis (TNA) is an assessment that allows an organisation to determine the skills gap (skills needs) in its employees, so that appropriate training opportunities can be tailored to meet the needs of its employee to do their work better. Unfortunately, training provided without prior assessment of the needs of employees could result in training being provided to people who do not need it, or the wrong training being provided. TNA is required for employment purposes across most verticals (Gould D. et.al, 2004). It has recently been applied to cybersecurity related roles and including information technology to evaluate and design a systematic package of learning for IT staff. TNA should be an ongoing continuous process useful for determining training needs. This has become even more important in cybersecurity related roles, first, because of the skills gap in cybersecurity roles, and secondly, and more fundamental, because of the increasing change and the introduction of new aspects in cybersecurity. For example, there are emerging aspects in cybersecurity that were never thought about some years ago, such as machine learning in cybersecurity, digital forensics, security analytics, cybersecurity operations centre, Blockchain in cybersecurity etc. These advances in cyber have brought its own challenges, and these may have exacerbated the already known cyber skills shortage. Due to these reasons, cybersecurity demands a coherent and structured learning and training programme. According to (Brown J., 2002), TNA is required to determine and create an effective training programme. We argue that to create an efficient cybersecurity programme to support Cyber Foresight, a continuous training needs assessment should be in place, and a structured training programme that stems from TNA must be deployed for the organisation. Through providing appropriate training to operator (e.g., security analysts, administrators and incident responders) will their understanding of situations improve, and arguably, combined with a healthy mental wellbeing should improve their overall situational awareness.

- **KSA (Knowledge, Skills and Abilities)**

The human operators should gain the necessary and appropriate knowledge, skills and abilities required to perform their duties regardless of the domain.

*Knowledge* is gained through training, reading, education and apprenticeship. A cybersecurity analyst, like a medical doctor, a nurse or an engineer needs to be taught the basics and the foundation of their profession, this may be through attending college, university or self-schooling, whichever, they need the knowledge required for the professional work they intend to undertake.

*Skills* are gained once knowledge is achieved. Skill is gained by the constant application of knowledge. For example, a Cyber security graduate is taught the foundation of computing, programming, information security and networking at college, university or tertiary institutions of some sort; but that candidate once graduated will need to apply the knowledge gained from education at work and by constantly doing so, he/she will perfect and gain the skills required to become a skilled worker in their various specialties. Skill is a specific aspect of knowledge. Knowledge is broad while skill is narrower and more focused. For example, one may have knowledge about computing but lack skills to repair computers. A typical example is a candidate who earns a bachelor's degree in computing, may not have the skills to programming and code development but may have the skills to repair computers; therefore, although he/she is knowledgeable in computing but lacks programming skills.

*Ability* is a higher level of skill. We define ability as the level of applicability of a specific skill. Ability levels are used to distinguish and differentiate expertise and competency levels. Frameworks exist such as the SFIA Framework (SFIA, 2018) that defines different levels of ability and competency levels (a.k.a. Responsibility levels) per specialist areas, such as Practitioner level, Senior Level, and Lead Level. Each level reflects the ability, competency, and responsibility required to appropriately execute that role and the responsibility accorded to that level, e.g., degree of autonomy, influence, complexity, knowledge and business skills. Hence, ability is not just about skills but also, includes required responsibility.

To gain Cyber Foresight, human operators should possess the requisite KSA appropriate for their role, responsibility and domain. This is more important these days than ever because of the changing nature of our modern technologically driven world. For example, computers are far more automated than ever and the advent of big data, cloud computing and the use of machine learning and artificial intelligence means that the operator is overwhelmed with far more information than were previously seen. Plus, these sets of information are provided at pace never seen before. Operator awareness, knowledge and skills are challenged every day and the level of sophistication and advancement means that operators must continuously train, reskill and upskill. Also, the advancement means that techniques used, say in cybersecurity to attack systems and to compromise critical assets are far more advanced these days. Emerging threats and bad actors use obscure programming techniques to evade detection (Computer Weekly, 2022), while others use a combination of automation, bots etc. to evade detection, either by conducting activities at pace that existing defences find challenging to detect or are overwhelmed that they may fail to spot or detect the attack. Hence, the knowledge, skills and abilities required to detect modern complex cyber-attacks do require a multidisciplinary approach. That is, a multidisciplinary knowledge, skills and abilities ranging from linguistics, psychology, human factors to cybersecurity. The threats facing modern society is varying, complex and sophisticated, as such, the workforce requires multidimensional and multidisciplinary approach.

### C4.2. Human Factors Blockers

Human Factors Blockers are factors, circumstances or preconditions that inhibit the performance, cognition, and overall mental wellbeing of an operator or team / group of operators. We discuss Human Factor Blockers from a standpoint of the cyber operators rather than a generalist contribution, therefore, we note that the list of Human Factor Blockers discussed in this paper is neither aimed to be exhaustive nor complete.

- **Comprehension**

A cyber programme, (e.g., a SOC) is one that comprises hundreds of thousands of assets (e.g., endpoints, servers, switches, routers, etc.), tens of different and diverse security controls (e.g., IDS, firewalls, anti-malware gateways, anti-virus, DLP, encryptors, key materials etc.), and several network infrastructures and segments which SOC operators continuously and protectively monitor. Each of the endpoints produce logs, events, flows, metrics and other business logic that are often different, use different formats and standards. While tools such as security information and event management (SIEM) may be used to normalise and correlate information, still there are myriad of intelligence, cues and prompts that are produced that the SOC analyst should understand, comprehend and therefore respond to. Maintaining a high-level situational awareness over such an environment is not trivial (Onwubiko, C. and Owens J.T., 2011). Endpoints usually produce and generate high volumes of logs; unfortunately, high volumes of logs do not equate to high fidelity in the evidence generated by the logs. Secondly, sensors are used to monitor, and correlate logs generated by the endpoints to determine whether they are genuine indicators of compromise. Unfortunately, when using multiple sensors (multisensor fusion), we are often faced with the challenge that evidence from multiple sensors can be complementary, conflicting and incomplete. The challenge lies in an approach to manage conflicting and incomplete evidence received from multiple sensors, and the reliability of such information or evidence. The evidence (in the form of logs, events or messages) from sensors are sometimes non-existent, conflicting or incomplete. That notwithstanding, the analyst is expected to understand such situations, and more so, are expected to investigate and respond to incidents that result from the myriad of disparate assets. The SOC analysts may be trained on one or many technologies, may understand the behaviour of one or many controls, but the symptomatic evidence that each produces under different situations are often different, and there will be situations that the analyst may have not experienced before, and even if they do, the incidents may be subtly different. We argue that analyst's situational awareness of cyber situation (e.g., data breach, cyber incident, etc.) is considerably good when it leverages tools and technologies that can detect many cyber situations; however, understanding, and knowing what the incidents are related to, and what may have caused the incidents are a totally different issue, and are by far beyond the basic cues and prompts offered by technology. The cyber analyst's mental ability, experience, cognition and understanding are all tested in this situation. Comprehension is not merely being aware of a situation, but rather, a much deeper understanding, insight and knowing (knowledge) far beyond the superficiality provided by the cues and alerts.

- **Mental Health & Wellbeing**

Our focus is on the cyber operator mental health and wellbeing, either as an individual, and as a team or group of operators; so, through this work, we assess factors that are likely to impact the mental health of cyber operators and consequently impact their cognition, productivity and efficiency.

  - Stress

Stress is a manifestation of poor health and sociological wellness. Symptoms of stress include lack of coordination, irritability, poor outlook, poor cognition and depression. Stress could be caused by many things, for example, *workplace related stress, workload related stress* and *technostress*. Other causes of stress include family-related stress, relationship-related stress, event-related stress etc. Our focus in this paper is stress related to operators at workplace, hence the three types of stress considered are: workplace related stress, workload related stress and technostress. Stress can lead to diminished responsibility, lack of awareness and a general poor outlook. It is imperative that cyber operators, especially those tasked with monitoring critical national, and business critical services are regularly assessed to ensure they are in good health conditions. Technostress is defined as psychosomatic illness caused by working with computer technology daily, or the negative psychological link between people and the introduction of new technologies (Wikipedia 2019). The fast-paced workplace, and the continuous changing technological landscape at workplace is shown to cause technostress (Ayyagari R. et al., 2011, Tarafdar, M., Tu, Q., and Ragu-Nathan, T.S., 2010, Lundgren M., and Bergstrom, E., 2019). Workplace-related stress and technostress should be minimised, and this can be achieved by ensuring that operators are not overwhelmed with endless activities and that there is a work life balance. Operators' workload should be regularly assessed as workload is known to contribute to stress, anxiety and causes operators to feel overwhelmed. Operators should be encouraged to have periodic time off from work, this is particularly essential for operators who work shifts, weekends, antisocial and long hours. This category of operators should be encouraged to take regular time off. Studies have shown that fatigue is a contributing factor to technostress, in addition to the rapid and constant changes in technology, pace of change and complexity (Ayyagari R. et al., 2011). It is therefore imperative that operators have appropriate rest and relaxation. We argue that technostress and work-related anxieties are not caused by workload alone but by the increasing complexity in technology, exacerbated by the pace of change and constant changes in organisational strategy. For example, in a certain government department, we observed that over a period of three months, more than 15 people resigned, majority of them cited reasons for their resignation to be work stress either resulting from workload and lack of clear directions by the senior management team or too many changes too frequently. While this is not a purely academic and scientific assessment, it is equally not a pure coincidence. People get disillusioned when there are far too frequent changes, especially when such changes do not seem to be accompanied by a clear sense of direction or purpose. Communication at workplace is lacking. People are not often informed on what is going on at workplace, or what may be coming their way, and so often, when communications are sent out, they are buried with other ephemeral emails that people do not bother reading. Email communications are often not clear and follow up conference calls or what is expected of employees are often missing or not properly articulated, and the rationale for such changes are not stated in such emails.

- ○ Anxiety

Anxiety, which can be caused by many things including fear, apprehension and uncertainty causes restlessness to operators which in turn may lead to poor cognition and wellbeing. Security-related-anxiety can be caused my many factors, such as a demanding job, pace of change in technology, especially those that require regular or uphill training and examination and, in some cases, require specialist prerequisite skills which many operators may not have.

- **Workload**

According to Lee C. et al. 2016, the two major causes of information security related stress are technology and job, and it has been shown that work overload and invasion of privacy are the main culprits (Ayyagari et al., 2011, Moore 2000). Studies have shown that workload, especially extreme and constant workload, has a correlation to poor cognition. It is no wonder why most safety related professions have restrictions and regulations to the number of hours they work. There is a federal regulation to the maximum number of hours pilots are restricted to fly per month. According to the Federation Aviation Administration (FAA, 2010), it is recognised that adequate rest of pilot is important to maintaining aviation safety. It is cited and shown that pilot's fatigue is a contributing factor and threat to flight safety (NYT, 1988). Similarly, clinicians, such as nurses and doctors are regulated, too. Unlike cyber operators, there is no such regulation that we know of, and because of this lack of regulation, cyber operators may work longer hours, especially during a major cyber incident or crisis. Further, cyber operators often work shifts, for instance in SOCs and NOCs, working weekends, and including anti-social hours and festive period. The backdrop of this is that where appropriate measures are nonexistent, cyber operators are overworked and this could lead to reduced cognition, poor mental agility and declined productivity.

- **Working Conditions & Practices**

Time is of the essence. Cyber incidents require immediate attention to respond to the attack, and most importantly to restore services. Cyber dictates the shortest mean time to detect and respond to a cyber-attack. This means, unlike traditional security incidents, such as physical break-in, data centre theft, etc., cyber incident dictates speed and accuracy. The reason is because if a cyber incident is left unattended, the attacker may delete, steal, tamper and destroy evidence that could be used to establish who the attacker was, how the incident happened, and what may have been compromised in addition. The cyber environment is always dynamic, continuous, and fast paced. Cyber incidents are sudden, ad hoc and unannounced, hence cyber operators must always be ready and available (a.k.a. cyber readiness). Like a crisis, cyber incidents require immediate response. Cyber operators are often overworked, under constant pressure, which is demanded of the job, especially at times of cyber incidents or crisis. With the increasing rise in cyber-attacks, cyber incidents and security breaches are also frequent. Emerging cyber-attacks are complex and sometimes coordinated, making them harder to detect. The complexity of emerging cyber-attacks mean that not only are the skills and experience of cyber operators challenged, but also, their aptitude and determination are equally tested. Those working in highly sensitive and high demanding environments and especially those protecting critical national security may be worse off. Further, the scarcity of good cyber experts a reflection of cyber skills shortage does not help the current situation. Working in an environment that is constantly demanding, while at the same time, you are required to react and swiftly respond in the event of an incident is challenging.

The rapidly changing technology and process mean that cognition and mental wellbeing of cyber operators may be negatively impacted. There is a high likelihood that over time cyber operators may experience poor cognition, reduced mental well-being and overall poor mental health resulting from working conditions and practices.

## IV. Cross-Domain Cybersecurity Situational Awareness

Situational awareness allows us to make sense or understand the cues that are produced by networks, systems and operational activities from which to gain better comprehension of the nature, state and impending situation. Business risks need understanding and so are business requirements and both go hand in hand. Understanding of the changing political, cultural and economic landscapes are important, too. Better decision making is achieved when all the disparate information is collated, correlated and analysed. While single events or events produced by one tool may be limited in view, we argue that having wide view of all the events produced across the estate allow for better situational awareness. Fig. 6 is an illustration of how Cyber Foresight can be gained on a single domain, e.g., Physical domain. A single domain spans multiple layers, such as physical layer, network layer, data layer, information layer, application layer and service layer. The awareness, comprehension, resolution and prediction still happen in a single domain. While foresight can be gained, but it is still of a single domain. There is no doubt that insights gained on a single domain but from various layers offer richer and better insights than that gained from a single layer on a single domain. We argue that to gain strategic Cyber Foresight and Cyber Superiority, insights from all the six domains should be considered and harnessed.

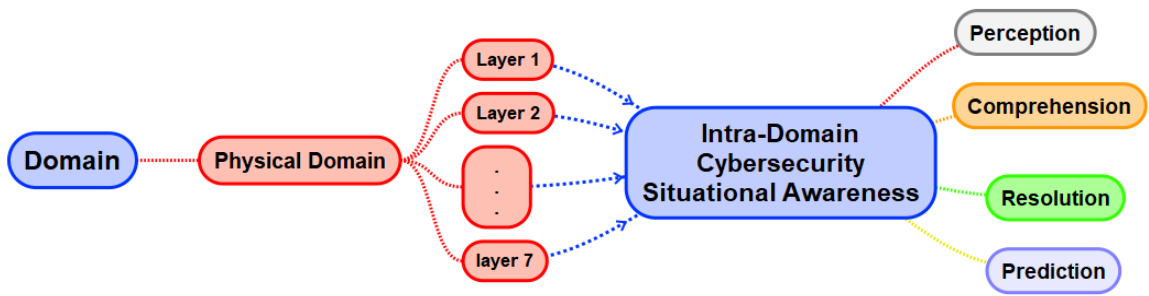

Figure 6: Intra-Domain Cybersecurity Situational Awareness (Intra-Domain CyberSA)

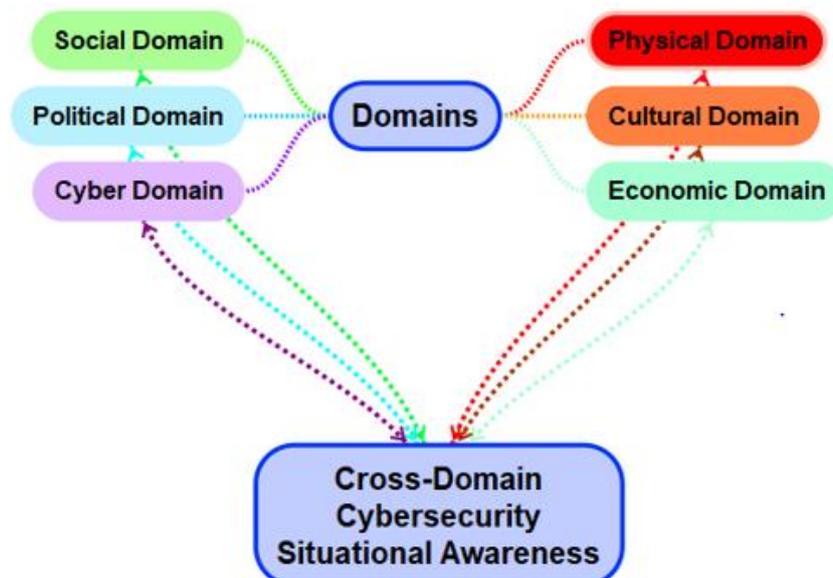

Figure 7: Cross-Domain Cybersecurity Situational Awareness (Cross-Domain CyberSA)

A multidimensional approach allows for information, knowledge, cues and intelligence gathered from the various domains to be collected, collated, processed, analysed and disseminated allowing the mission or organisation an uninterrupted access to a great deal of intelligence, which invariably offer enhanced situational awareness over intelligence or information from a singular, non-coordinated domain. In addition, important decision makings can be collaborative, and shared across the domains in a coordinated fashion allowing for cooperation and accountability.

It is of note to distinguish between intra-domain (see Fig. 6) and inter-domain (see Fig. 7) cybersecurity situational awareness.

*Intra-domain cybersecurity situational awareness* (Intra-domain CyberSA) is the collection, processing and analysis of cues, prompts, information and intelligence from multiple layers of a single domain (see Fig. 6). For example, the collection, processing and analysis of cues, information and intelligence in, and of a single domain e.g., Physical domain.

*Inter-domain cybersecurity situational awareness* (Inter-domain CyberSA) is the collection, processing and analysis of cues, prompts, information and intelligence from multiple layers across multiple domains (see Fig. 7). For example, collection, processing and analysis of cues, prompts, information and intelligence across the six domains of the Physical, Cultural, Economic, Social, Political and Cyber. We argue that inter-domain situational awareness offers strategic Cyber Foresight over situational awareness gained of a single domain. The rationale is that awareness, comprehension and prediction stemming from processing of cues and prompts from a single domain is limited and does not offer the overarching understanding of impending issues in the other domains, neither is it aware of contextual information and intelligence outside its own domain. Consequently, which may impact the accuracy, reliability and completeness needed, and therefore is likely not to provide a complete picture of the entire situation.

Table 4: Differences between intra-domain and inter-domain relationship

| Intra-domain | Inter-domain |
|---|---|
| Within a single domain | Across multiple domains |
| Across multiple layers in a single domain | Across multiple layers across multiple domains |
| Localised situational awareness of a single domain | Domain-wide situational awareness across multiple domains and across multiple layers |
| Cyber Foresight of a single domain | Cyber Foresight across multiple domains offering holistic and strategic foresight |
| Constrained and limited | Scalable and robust |

## V. Conclusions and Future work

Cybersecurity (i.e., secure and safe cyberspace) should be everyone's goal and priority – citizens, industry, academia, government and nations. It should also be a multidisciplinary and multidimensional endeavour requiring concerted efforts from everyone – lawmakers, law enforcement, government, policy makers, investors, practitioners, technologists, stakeholders and supply chain. Cooperation through multilateral agreements, open and transparent voluntary commitments – rules of engagement and responsible behaviour in cyberspace should be the norm.

Cyber Foresight – the ability to understand current cyber situations or circumstances, the preparation and capability to identify future cyber situations, and the capacity to address these situations – is gained through a multidisciplinary and multidimensional continuous approach. Cyber Foresight encompasses knowledge, understanding, and insight into the current cyber situations or circumstances, the readiness and preparedness to understand plausible future situations and/or scenarios, and the capacity to address (deter, deflect and resist) the emerging future cyber situations. As noted by the European Union Agency for Cybersecurity (ENISA, 2021), gaining or acquiring Cybersecurity Foresight is complex. It requires multi-stage and continuous endeavour that involves strategic planning, preparation and development aimed for informed decision marking for securing current and future possible developments. So, the goal for National Cyber Foresight should focus on:
- responsible behaviour in cyberspace,
- capability enhancement to gain understanding and insights of 'bad actors' and noncooperating or hostile Nations,
- cooperation among multilateral stakeholders, and
- through a multidisciplinary and multidimensional collaboration to understand plausible emerging and future cyber situations, and
- capacity to deter, deflect and resist adversarial infringement.

Planning, designing, deploying and executing cybersecurity programmes (e.g., cybersecurity capacity centres, National cybersecurity agency, cyber fusion centres etc.) requires a multidisciplinary and multidimensional approach. As discussed in this paper, multidimensional cybersecurity fosters collaboration and cooperation across multiple domains, not just one domain/dimensional approach of focusing on technical capabilities alone. An adoption of an inclusive and encompassing approach considering impacts from and to the other domains, e.g., Physical, Cultural, Economic, Social, Political and Cyber domains.

Business, Operational, Technological and Human factors contribute to achieving multidimensional cybersecurity. Business Factors allow organisation-wide business requirements to be successfully executed toward achieving cost, performance and business efficiencies. Operational Factors ensure operational aspects of the cybersecurity programme are addressed, and they help fulfill the objectives of the Technology Factors when implemented correctly and also, ensure Business Factors are met.

Human Factors in cybersecurity cannot be overemphasised. Strategic cybersecurity foresight is not a product, or an end game but a journey. It is underpinned on our determination to embed and practice consciously the rules of responsible behaviour to improve the quality of our service, diversity, equity and inclusion, consideration for ESG and sustainability, a drive to ensure ethical use of artificial intelligence and cognition for adequacy in our data privacy practices.

Situational awareness is a human attribute; therefore, it is so challenging to demonstrate when it is earned from a technology standpoint. Several factors constituting technology can enable foresight to be gained by the operators of the systems through training, experience, training and needs assessment (TNA), knowledge, skills and abilities. Unfortunately, there are other factors that could inhibit or limit operator situational awareness such as stress, anxiety, workload, environment and cognition.

Future work should explore a multidisciplinary collaboration of domain experts across Foresight, Linguistics, Humanities, Psychology, Sociology, Cybersecurity, Artificial intelligence and Policy to develop sample application use cases for Cyber Foresight for selected scenarios or cyber situations. The selected cyber scenarios should focus on generally applicable scenarios initially.

75Computer Weekly (2022), "Malicious Actors turn to Obscure Programming Languages". 26 July 2021. Retrieved Jan 2022 from https://www.computerweekly.com/news/252504470/Malicious-actors-turn-to-obscure-programming-languages
CNN (2020), "2016 Presidential Campaign Hacking Fast Facts", CNN Editorial Research, Updated 1512 GMT, October 28, 2020. Retrieved August 2021 from https://edition.cnn.com/2016/12/26/us/2016-presidential-campaign-hacking-fast-facts/index.html
Cumiford D. L. (2006), "Situation Awareness for Cyber Defense", 2006 CCRTS – The State of the Art and the State of the Practice, Sandia National Laboratories, MS 0455, USA, 2006
D'Amico A. and Kocka M., (2005), "Information Assurance Visualisation for Specific Stages of Situational Awareness and Intended Uses: Lessons Learned", Workshop on Visualisation for Computer Security, USA, 2005.
Dawis R. and Lifquist L. H. (1987), "Measurement of person-environment fit and prediction of satisfaction in the theory of work adjustment", J Vocat Behav 1987; 31(3):297-318
Dictionary Online (2019), "Foresight". https://www.dictionary.com/browse/foresight
Dictionary of Military and Associated Terms (2005). Retrieved April 8, 2019, from https://www.thefreedictionary.com/operational+architecture
DoD MNE7 (2013), "Multinational Experiment 7, Outcome 3 – Cyber Domain, Objective 3.2 - Information Sharing Framework, 22 January 2013. Accessed 22nd Jan 2019 https://www.nist.gov/sites/default/files/documents/2017/04/26/dod_js_j7_part_2_022713.pdf
Duerst M. and Eichhorn E. (2020) "The Future State of Value-Driven Consumer Goods", Gartner
Endsley M. R. (1995), "Toward a Theory of Situation Awareness in Dynamic Systems. Human Factors Journal 37(1), 32-64, March 1995.
Endsley M. R., and Connors E., (2012), "Situation Awareness in Cyber Operations", in Perspectives on the Role of Cognition in Cyber Security" in the Proceedings of the Human Factors and Ergonomics Society 56th Annual Meeting, 2012, page 268-271.
Endsley M. R. and Garland D. J., (Eds) (2000), Situation Awareness Analysis and Measurement; Mahwah, NJ; Lawrence Erlbaum Associates, ISBN: 0-8058-2133-3, 2000.
Eiza M. H. (2017), "Application of Cyber Situational Awareness and Cyber Security in Connected Vehicles", Cyber Science 2027, London, UK
ENISA (2021), "Foresight Challenges – A study to enable foresight on emerging and future cybersecurity challenges". European Union Agency for Cybersecurity, ENISA. November 2021. Retrieved Feb 2022 from https://www.enisa.europa.eu/publications/foresight-challenges
FAA, (2010) Fact Sheet – Pilot Flight Time, Rest, and Fatigue, January 27, 2010 - https://www.faa.gov/news/fact_sheets/news_story.cfm?newsId=6762
Finland Cybersecurity Strategy, (2019), Finland´s Cyber security Strategy, 2019. Retrieved August 2021 from https://turvallisuuskomitea.fi/wp-content/uploads/2019/10/Kyberturvallisuusstrategia_A4_ENG_WEB_031019.pdf
Frangopoulos E.D., Eloff M.M., Venter L.M. (2008) "Social Aspects of Information Security", Proceedings of the ISSA 2008 Innovative Minds Conference, ISSA 2008, Gauteng Region (Johannesburg), South Africa, 7-9 July 2008
Gandhi R., Sharma A., Mahoney W., Sousan W., Zhu Q. and Laplante P. (2011), "Dimensions of Cyber-Attacks, Social, Political, Economic, and Cultural", IEEE Technology and Society Magazine, Spring, 2011
Gartner (2021), "ESG Risk by the Numbers: Benchmarking ESG Disclosures". Published 4 August 2021 – ID G00756331 by Enterprise Risk Management Research Team.
Gould D., Kelly D., White I. and Chidgey J., (2004), "Training Needs Analysis. A Literature Review and reappraisal", Science Direct, International Journal of Nursing Studies 41 (2004) 471-486
Gregoire M. and Beaudoin L., (2005), "Visualisation for Network Situational Awareness in Computer Network Defence", Proceedings of the *Visualisation and the Common Operational Picture, pp. 20-1-20-6, RTO-MP-IST-043*, 2005.
Hall D. L., and McMullen S. A. H., (2004), "Mathematical Techniques in Multisensor Data Fusion", 2nd Edition, ISBN: 1-58053-335-3, 2004
Hall M. J., Hansen D. D., Jones K. (2015), "Cross-domain situational awareness and collaborative working for cyber security", in the 2015 IEEE International Conference on Cyber Situational Awareness, Data Analytics and Assessment (CyberSA), 2015.
Ikwu R. (2017), "Multi-Dimensional Structural Data Integration for Proactive Cyber-Defense", published in the IEEE International Conference on Cyber Situational Awareness, Data Analytics and Assessment (Cyber SA, 2017), 2017
Koh D. (2020), "The Geopolitics of Cybersecurity – Cooperation among states must underpin efforts to create a safer, more secure, and interoperable cyberspace" December 09, 2020. The Diplomat. Retrieved August 2021 from https://thediplomat.com/2020/12/the-geopolitics-of-cybersecurity/
Kotler, P & Armstrong, G., (2004). "Principles of Marketing", Upper Saddle River, New Jersey: Prentice Hall, 2004
Lesk M. (2011), "Cybersecurity and Economics", *IEEE Security & Privacy Economics*, pp. 76-79, November/December 2011
Lipp M., Schwarz M., Gruss D., Prescher T., Haas W., Fogh A., Horn J., Mangard S., Kocher P., Genkin D., Yarom Y. and Hamburg M. (2018), "Meltdown: Reading Kernel Memory from User Space", in the 27th USENIX Security Symposium, USENIX Security 2018
Lockheed Martin Cyber Kill Chain (2016) - Cyber Kill Chain, Accessed 8th April 2019, https://www.lockheedmartin.com/en-us/capabilities/cyber/cyber-kill-chain.html
Lundgren, M., and Bergstrom, E., (2019), "Security-Related Stress: A Perspective on Information Security Risk Management", in the proceedings of the IEEE International Conference on Cyber Security and the Protection of Digital Services (Cyber Security 2019), Oxford, June 3-4, 2019
Marotta A., and Dr. Keri Pearlson K., (2019) A Culture of Cybersecurity at Banca Popolare di Sondrio, the Twenty-fifth Americas Conference on Information Systems, Cancun, 2019
Midler M. (2020), "Ransomware as a Services (RaaS), Software Engineering Institute (SEI), Carnegie Mellon University, 2020. Retrieved Aug 2021 from https://insights.sei.cmu.edu/blog/ransomware-as-a-service-raas-threats/
Mitre ATT&CK, (2017) - Mitre Att&ck framework - https://attack.mitre.org/
Nakhla N., Perrett K. and McKenzie C. (2017), "Automated Computer Network Defence using ARMOUR: Mission-oriented decision support and vulnerability mitigation, published in the IEEE International Conference on Cyber Situational Awareness, Data Analytics and Assessment (Cyber SA, 2017), 2017.
NATO Cooperative Cyber Defence Centre of Excellence (CCD COE). 2015
New York Times, (1988) - Pilots' Fatigue Termed Threat To Safe Flying. Accessed 7th April 2019 https://www.nytimes.com/1988/06/05/us/pilots-fatigue-termed-threat-to-safe-flying.html

## BIOGRAPHICAL NOTES

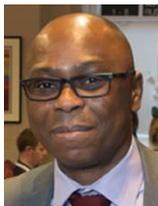

**Dr Cyril Onwubiko** is a Senior Member of the IEEE, Distinguished Speaker and Board of Governor, IEEE Computer Society. He's a Senior Director, Enterprise Security Architecture at Pearson – *the world's learning organisation*, and Director, AI, Blockchain & Cyber Security at Research Series Limited, where he is responsible for directing AI strategy, Blockchain governance and Cyber security. He is a leading scholar in Cyber Situational Awareness (Cyber SA), Cyber Security, Data Fusion, Intrusion Detection Systems and Computer Network Security; and vastly knowledgeable in Information Assurance, HMG Security Policy Framework (SPF) and Risk Assessment & Management. Cyril holds a PhD in Computer Network Security from Kingston University, London, UK; MSc in Internet Engineering, from University of East London, London, UK, and BSc, first class honours, in Computer Science & Mathematics. He has authored several books including "Security Framework for Attack Detection in Computer Networks" and "Concepts in Numerical Methods." He co-edited the book on "Situational Awareness in Computer Network Defense: Principles, Methods & Applications" published by IGI Global, USA. Cyril is the Editor-in-Chief of the International Journal on Cyber Situational Awareness (IJCSA), and founder of the Centre for Multidisciplinary Research, Innovation and Collaboration (C-MRiC), London, UK.

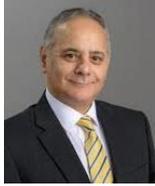

**Professor Karim Ouazzane** is a professor of Computing and Knowledge Exchange, Director of Research and Enterprise, Chair of the European Cyber Security Council (Brussels), and founder of the Cyber Security Research Centre (CSRC), London Metropolitan University, London, UK. His research interests include artificial intelligence (AI) applications, bimodal speech recognition for wireless devices, cyber security and big data, computer vision, hard and soft computing methods, flow control and metering, optical instrumentation and lasers. He has carried out research in collaboration with industry through a number of research schemes such as The Engineering and Physical Sciences Research Council (EPSRC), KTP, EU Tempus, LDA (London Development Agency), KC (Knowledge Connect) and more. He has also published over 100 papers, three chapters in books, is the author of three patents and has successfully supervised 13 PhDs. He is a member of the Oracle Corporation Advisory Panel.

# REFERENCE

**Reference to this paper should be made as follows**: Onwubiko C. and Ouazzane K. (2021). Multidisciplinary Cybersecurity Framework for Strategic Foresight. *International Journal on Cyber Situational Awareness*, Vol. 6, No. 1, pp46-77.